\documentclass[%
 reprint,
 amsmath,amssymb,
 aps,
]{revtex4-2}

\usepackage{xcolor}
\usepackage{amsfonts}
\usepackage{amsmath}
\usepackage{enumitem}
\usepackage{tikz}
\usepackage{wasysym}
\usepackage{hyperref}
\usetikzlibrary{quantikz}

\definecolor{darkgreen}{rgb}{0.13, 0.61, 0.13}

\newcommand{\oplusbar}{\overline{\oplus}}
\newcommand{\otimesbar}{\overline{\otimes}}
\newcommand{\norm}[1]{\left\lVert #1 \right\rVert}
\newcommand{\hsnorm}[1]{\left\lVert #1 \right\rVert_{\text{\tiny HS}}}

\DeclareMathOperator{\Tr}{\text{Tr}}

\newcommand{\figref}[1]{Figure~#1}
\newcommand{\tabref}[1]{Table~#1}
\newcommand{\secref}[1]{Section~#1}
\newcommand{\appref}[1]{Appendix~#1}

\renewcommand\Re{\operatorname{Re}}
\renewcommand\Im{\operatorname{Im}}

\definecolor{softblue}{rgb}{0.85, 0.90, 0.98}
\definecolor{softred}{rgb}{0.97, 0.81, 0.8}
\definecolor{softgreen}{rgb}{0.83, 0.91, 0.83}
\definecolor{softpurple}{rgb}{0.88, 0.83, 0.90}
\definecolor{softorange}{rgb}{1.0, 0.90, 0.8}

\begin{document}

\title{Trotterless Simulation of Open Quantum Systems for NISQ Quantum Devices}

\author{Colin Burdine}
\email{Colin\_Burdine1@baylor.edu}
\author{Enrique P. Blair}%
\affiliation{%
 Department of Electrical and Computer Engineering, Baylor University, Waco, Texas 76798-7316, USA
}%

\begin{abstract}
The simulation of quantum systems is one of the flagship applications of near-term NISQ (noisy intermediate-scale quantum) computing devices. Efficiently simulating the rich, non-unitary dynamics of open quantum systems remains challenging on NISQ hardware. Current simulation methods for open quantum systems employ time-stepped Trotter product formulas (``Trotterization") which can scale poorly with respect to the simulation time and system dimension. Here, we propose a new simulation method based on the derivation of a Kraus operator series representation of the system. We identify a class of open quantum systems for which this method produces circuits of time-independent depth, which may serve as a desirable alternative to Trotterization, especially on NISQ devices.
\end{abstract}

\maketitle

\tableofcontents

\section{Introduction}
\label{sec:introduction}

The quantum computer is an emerging technology with the potential to greatly accelerate progress in the fields of chemistry, physics, and materials science \cite{preskill_quantum_2018, daley_practical_2022}. One of the most extensively researched near-term applications of quantum computing is the simulation of physical systems, since state-of-the art NISQ (noisy intermediate-scale quantum) hardware has enabled accurate simulations of various kinds of systems \cite{clinton_hamiltonian_2021}. Recently, significant effort has been devoted toward the representation and simulation of closed many-body quantum systems, such as spin-lattice systems \cite{smith_simulating_2019}, fermionic systems \cite{jiang_quantum_2018, kitaev_quantum_1997}, and bosonic systems \cite{lamata_efficient_2014, macridin_digital_2018}. However, relatively few studies have investigated the time evolution of open quantum systems, which exhibit complex non-unitary dynamics due to interactions with an environment. Since quantum computation is inherently unitary, efficiently simulating the non-unitary dynamics of open systems on quantum computers remains a challenging yet important problem, especially on state-of-the art NISQ hardware \cite{tacchino_quantum_2020}.

\subsection{Related Work}
\label{sec:related_work}

Many different approaches have been taken to simulate open quantum systems, such as explicitly modeling the environment \cite{wang_simulating_2023, su_quantum_2020}, numerically evolving the non-unitary dynamics of the system \cite{muller_engineered_2012, kamakari_digital_2022, kliesch_dissipative_2011, schlimgen_quantum_2022, jo_simulating_2022, cygorek_simulation_2022, del_re_driven-dissipative_2020}, harnessing device noise to model the environment \cite{guimaraes_noise-assisted_2023}, or representing the system's dynamics in closed form, either as an arbitrary quantum channel \cite{david_digital_2024} or a Kraus representation \cite{hu_quantum_2020, dive_quantum_2015, garcia-perez_ibm_2020}. Variational approximations of the steady states of open systems have also been successfully demonstrated \cite{endo_variational_2020}. In a majority of these studies \cite{guimaraes_noise-assisted_2023, kamakari_digital_2022, kliesch_dissipative_2011, schlimgen_quantum_2022, jo_simulating_2022, cygorek_simulation_2022, del_re_driven-dissipative_2020, hu_quantum_2020, dive_quantum_2015, garcia-perez_ibm_2020}, systems are considered under the Born-Markov approximation, in which environmental coupling is weak and correlations are assumed to decay rapidly such that the system exhibits Markovian dynamics (i.e. the environment is effectively ``memoryless"). The dynamics of Markovian systems are governed by one of a number of master equations, such as the stochastic Schr\"odinger equation \cite{jacobs_quantum_2014} or the Lindblad equation \cite{breuer_theory_2002}. The most common and readily generalizable approach to simulating Markovian systems is the use of Suzuki-Trotter product formulas (``Trotterization") \cite{trotter_product_1959, hatano_finding_2005}. This approach uses discrete-time numerical methods to compute the time evolution of sparse non-commuting components of the Hamiltonian and any non-Hermitian operators that model environmental interactions \cite{berry_efficient_2007,kamakari_digital_2022}. However, Suzuki-Trotter-based methods require quantum circuits that increase linearly with the time evolution period and quadratically with the system dimension \cite{kamakari_digital_2022, kliesch_dissipative_2011, childs_theory_2021}. This is problematic, especially if these methods are to be realized on NISQ devices, where factors such as device topology and noise must be taken into account \cite{clinton_hamiltonian_2021}.

Motivated by the constraints of modern quantum hardware, significant strides have been made toward finding representations of open quantum systems that either allow for quantum resource-efficient Trotterization or avoid Trotterization altogether. In a few cases, Trotterless methods for the time-evolution of open systems have been found \cite{hu_quantum_2020, dive_quantum_2015, garcia-perez_ibm_2020}, though these methods apply to only a limited number of toy example systems, where the dynamics admit a known closed-form Kraus representation. For arbitrary open quantum systems, computing minimal Kraus representations is known to be classically hard \cite{andersson_finding_2007}. Nonetheless, due to the inherent non-uniqueness of Kraus representations, it has remained an unexplored question whether there exist more general classes of systems that admit non-minimal or approximate Kraus representations. It also remains to be seen if such representations may facilitate new NISQ-friendly simulation methods.

\subsection{Main Results}
\label{sec:main_results}

\begin{figure}
    \centering
    \includegraphics[width=\linewidth]{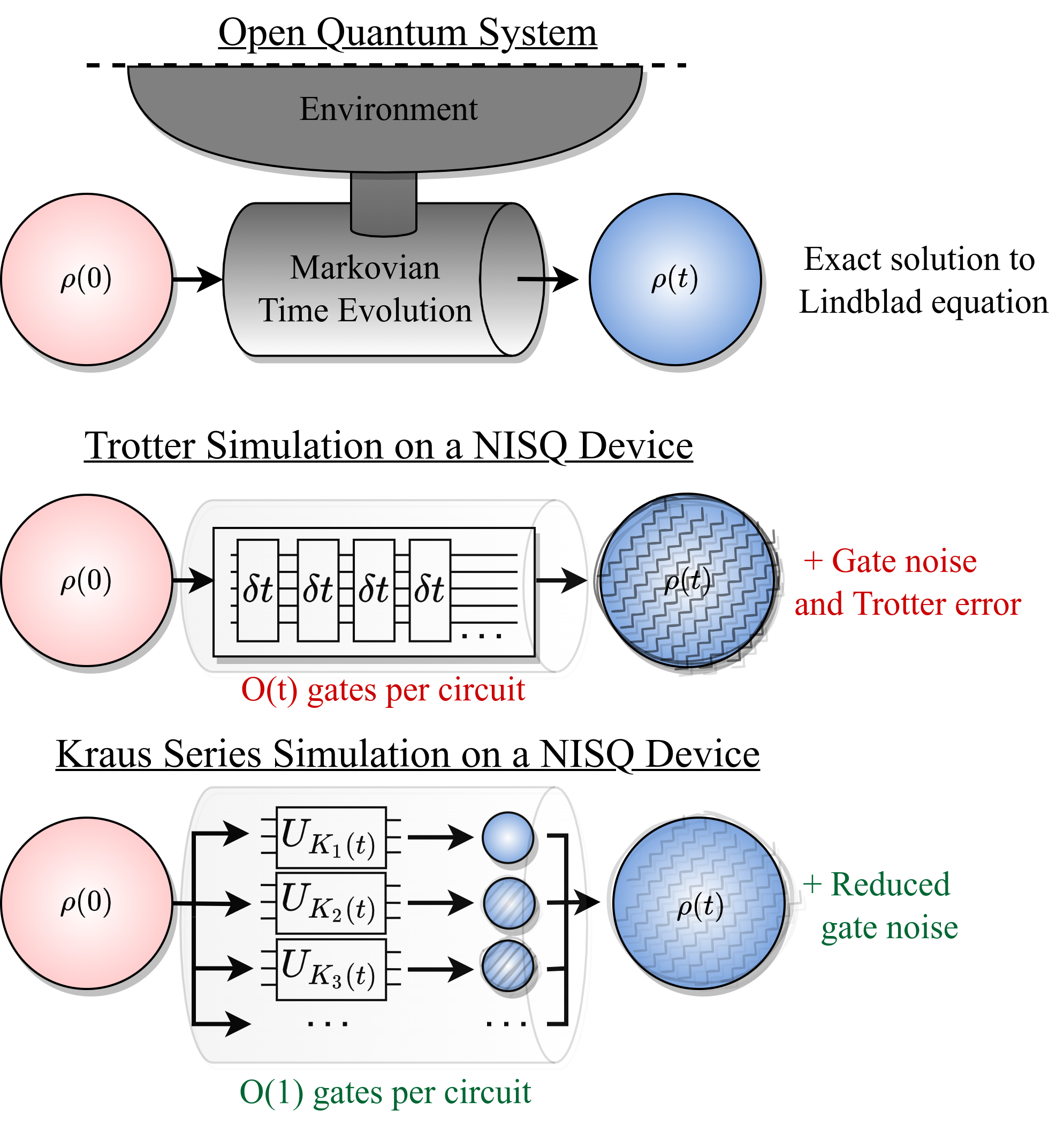}
    \caption{Overview of approaches to simulating Markovian open quantum systems. When Trotter-based methods are applied, deep circuits are required with $O(t)$ gate complexity. For complex systems, each Trotter step ($\delta t$) may require many gates to numerically approximate. On NISQ devices, this results in noise and Trotter approximation error. For certain kinds of systems, our Kraus Series approach distributes the complexity of the simulation over many short circuits, thereby reducing the effect of noise.}
    \label{fig:overview}
\end{figure}

In this paper, we make preliminary steps toward identifying and characterizing systems which admit ``Trotterless" representations. More specifically, we identify a general class of Born-Markov open quantum systems described by the Lindblad master equation that admit closed-form Kraus operator series representations. We show that the time evolution of these systems can be computed to a desired order of accuracy for any evolution period $t \ge 0$ using a series of parameterized quantum circuits (one per Kraus operator) with a gate complexity independent of $t$. The closed-form $t$-independent property of these circuits makes them ideal candidates for simulation on NISQ computers, since they do not require Trotterization or other quantum digital simulation methods. This is illustrated in \figref{\ref{fig:overview}}

To characterize the dynamics of Lindblad-type open quantum systems, we introduce the effective Hamiltonian superoperator $\mathcal{H}$ and Lindblad superoperator $\mathcal{L}$. We show that $\mathcal{H}$ and $\mathcal{L}$ can be used to identify this class of systems, provided that the following three general conditions are met:

\begin{enumerate}
    \item The system's effective Hamiltonian dynamics $\exp(t\mathcal{H})$ can be simulated efficiently via a quantum circuit with gate complexity independent of $t$.
    \item The system's Lindblad operators $L_n$ are sparse and can be implemented through a quantum circuit of bounded gate complexity.
    \item The system's $\mathcal{H}$ and $\mathcal{L}$ superoperators satisfy the commutation relation $[\mathcal{H},\mathcal{L}] = \alpha \mathcal{L} + c$ for real scalars $\alpha, c \ge 0$.
\end{enumerate}

As a more concrete case, we show that systems with a Hamiltonian $H$ and a set of sparse Lindblad operators $\{ L_n \}_{n=1}^{N_L}$ that satisfy the commutation relations

\begin{enumerate}[label=(\roman*)]
    \item $[H,L_n^\dagger L_n] = 0$ (for all $L_n$)
    \item $[L_n^\dagger L_n, L_{n'}^\dagger L_{n'}] = 0$\quad (for all $L_n, L_{n'}$)
    \item $[H, L_n] = \nu L_n$\\[2mm] 
        (for some $\nu \in \mathbb{C}$ with $\Im[\nu] \ge 0$, for all $L_n$)
    \item $\sum_{n'} \gamma_{n'}[L_{n'}^\dagger L_{n'}, L_n] = \lambda L_n$\\[2mm] 
        (for some $\lambda \in \mathbb{C}$ with $\Re[\lambda] \le 0$, for all $L_n$)
\end{enumerate}

\noindent all satisfy conditions 1-3 above and can be simulated with a series of quantum circuits, each of which has time-independent gate complexity at most $O(d^2)$, where $d$ is the dimension of $H$. Although this series generally consists of an infinite number of circuits, we show that accurate approximations of the final state can be made if the series is truncated to a finite number of circuits, thereby allowing for a tunable trade-off between desired accuracy and number of circuits. We also show that for some systems, exact solutions can be obtained with a finite number of circuits.

Finally, we show that two widely studied open quantum systems, the multi-qubit continuous-time Pauli channel and the damped quantum harmonic oscillator, both satisfy conditions 1-3 and can be simulated efficiently with a series of ``Trotterless" quantum circuits with $t$-independent depth. Both of these systems play a crucial role in modeling the Markovian decoherence of spin-$1/2$ fermionic systems and bosonic systems respectively.

\section{Background}
\label{sec:background}

\subsection{Lindblad Equation}
\label{sec:lindblad_equation}

In this paper we consider non-relativistic open quantum systems under the Born-Markov approximation. The time evolution of these systems forms a dynamical semigroup, and the master equation that generates their non-unitary dynamics can be written in the Lindbladian (also known as the diagonalized GKSL) form \cite{breuer_theory_2002}:

\begin{equation}
\begin{aligned}
\dfrac{d\rho}{dt}(t) &= -\frac{i}{\hbar}[H, \rho(t)] \\
&\qquad + \sum_{n} \gamma_n \left( L_n \rho(t)L_n^\dagger - \frac{1}{2}\lbrace L_n^\dagger L_n, \rho(t) \rbrace \right).
\label{eqn:lindblad}
\end{aligned}
\end{equation}

Above, $H$ is the system Hamiltonian and the $L_n$ are Lindblad operators with corresponding damping parameters $\gamma_n \ge 0 $. These operators are often imposed phenomenologically on the system, since a large number of Lindblad operators (sometimes even an infinite number \cite{liu_kraus_2004}) are usually needed to model the Markovian dynamics of the total unitary evolution of a system and environment. For example, in systems with quadratic Hamiltonians expressed in terms of creation and annihilation operators $\hat{a}_n^\dagger, \hat{a}_n$, i.e:
\begin{equation}
H = \sum_{n} c_n\hat{a}_n^\dagger \hat{a}_n,
\label{eqn:quadratic_H}
\end{equation}
the lowering operators $\hat{a}_n$ are often used as the Lindblad operators $L_n$ with appropriate choice of damping $\gamma_n$ to model spontaneous emission and decay within the system. However, in the simulation of other systems, more complex Lindblad operators (e.g. depolarizing and dephasing operators) have been applied to model various forms of environmentally-induced decoherence. These kinds of Lindblad operators are seen in the continuous-time Pauli channel, which is used to model unbiased qubit decoherence in quantum information processors.

\subsection{Kraus Operators}
\label{sec:kraus_operators}

It can be shown that the time evolution of a system $\rho(t)$ satisfying \eqref{eqn:lindblad} generates a completely positive trace-preserving (CPTP) map $\mathcal{E}_t$, which forms a quantum channel for each time evolution period $t \ge 0$ \cite{breuer_theory_2002}. This channel can be represented non-uniquely as a sum of time-dependent Kraus operators:

\begin{equation}
\rho(t) = \mathcal{E}_t(\rho(0)) = \sum_{i} K_i(t)\rho(0)K_i^\dagger(t).
\label{eqn:krauss}
\end{equation}

The Kraus operators $K_i(t)$ must satisfy the constraint
\begin{equation}
    \sum_{i} K_i(t)K_i(t)^\dagger = 1
\end{equation}
for $t \ge 0$. Also, the time-dependence of the Kraus operators must be such that the time evolution map $\mathcal{E}_t$ satisfies the dynamical semigroup property
\begin{equation}
\mathcal{E}_{t_2}(\mathcal{E}_{t1}(\rho)) = \mathcal{E}_{(t_1+t_2)}(\rho),
\label{eqn:krauss_semigroup} 
\end{equation}
where $\mathcal{E}_{0} \equiv \mathcal{I}$ (the identity map).\\

\subsection{Superoperator Formalism}
\label{sec:superoperator_formalism}

Before proceeding, we must introduce the superoperator notation that we will use throughout the remainder of this paper. A superoperator is an operator that acts linearly on the space of operators. In this case, we will focus on superoperators that act linearly on the ``vectorized" density operator, $\vec{\rho}$, given by
\begin{equation}
    \vec{\rho} = \sum_{i,j} \bra{e_i}\rho\ket{e_j}(\ket{e_i} \otimes \ket{e_j}),
\end{equation}
where the kets $\ket{e_i}$ form a complete orthonormal basis. It can be shown that superoperators acting by a left and right operator of an $n \times n$ density matrix $\rho$ can be represented as an $n^2 \times n^2$ matrix equal to the Kronecker product of the left-acting operator and the transpose of the right-acting operator, which acts linearly on $\vec{\rho}$. For example, a superoperator $\mathcal{D}$ that acts on an operator $\rho$ through left and right multiplication by the respective matrices $A$ and $B^\dagger$ has the superoperator representation $(A \otimes \overline{B})$, where $\overline{B}$ denotes the complex conjugate of $B$. This gives rise to the following correspondence between operator and superoperator representations:
\begin{equation}
\mathcal{D}(\rho) = A \rho B^\dagger\ \Leftrightarrow\ \mathcal{D}\vec{\rho} = (A \otimes \overline{B})\vec{\rho}.
\label{eqn:superops}
\end{equation}
This correspondence is quite useful, since it lifts the dynamics of a system's density matrix into a larger space where the dynamics can be written as a matrix-vector product.

\section{Theory}
\label{sec:theory}

\begin{figure}
\centering
\includegraphics{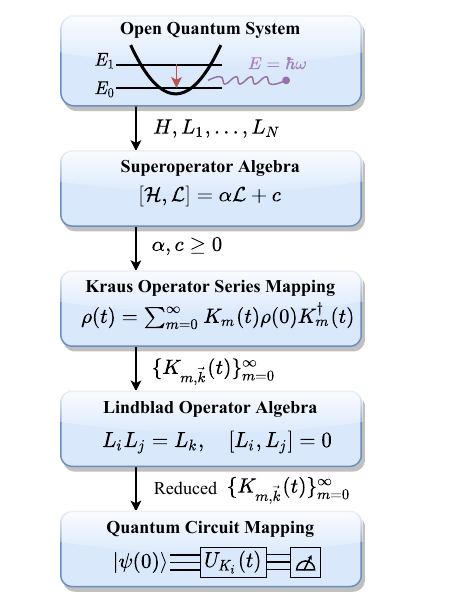}
    \caption{A summary of the steps required to realize open quantum systems satisfying conditions 1-3 as a series of quantum circuits. By exploiting the superoperator algebra, a Kraus series representation of a system can be found. This series can then be simplified using product and commutator identities of the Lindblad operator algebra (if any exist). Finally, each Kraus operator is mapped to a quantum circuit, which has depth independent of the time evolution period $t$.}
    \label{fig:theory_pipeline}
\end{figure}

In this section we will introduce our proposed framework for simulating open quantum systems that satisfy conditions 1-3. First, we will give an overview of our general framework for characterizing the dynamics of open systems in terms of the algebra of the effective Hamiltonian superoperator $\mathcal{H}$ and the Lindblad superoperator $\mathcal{L}$. Next, we will show how a Kraus series can be derived from $\mathcal{H}$ and $\mathcal{L}$ when the commutation relation in condition 3 is satisfied. We will also give examples of how useful identities in the Lindblad operator algebra can be applied to further reduce the complexity of the Kraus series. Finally, we will show how each Kraus operator in the series can be realized as a quantum circuit. A summary of this procedure is shown in \figref{\ref{fig:theory_pipeline}}. At the end of this section, we will also give more detailed analysis of the convergence of the Kraus series, and describe how the series can be truncated to a finite number of terms while still providing accurate results.

\subsection{General Framework}
\label{sec:general_framework}

We will begin by using the correspondence in \eqref{eqn:superops} to convert the right hand side of \eqref{eqn:lindblad} into superoperator form. To simplify this process, we first introduce the effective Hamiltonian operator $V_H$ and scaled Lindblad operators $V_n$: 
\begin{align}
    V_H &= H - \frac{i\hbar}{2}\sum_n \gamma_n L_n^\dagger L_n \label{eqn:V_H}\\
    V_n &= \sqrt{\gamma_n}L_n \label{eqn:V_n}
\end{align}

Next, we introduce the effective Hamiltonian superoperator $\mathcal{H}$ and Lindblad superoperator $\mathcal{L}$:
\begin{align}
    \mathcal{H} &= \left(\frac{-i}{\hbar} V_H\right)^{\oplusbar 2} \label{eqn:H_superoperator} \\[2mm]
    \mathcal{L} &= \sum_n \left(\sqrt{\gamma_n} L_n\right)^{\otimesbar 2} = \sum_n V_n^{\otimesbar 2} \label{eqn:L_superoperator}
\end{align}

In \eqref{eqn:H_superoperator}, \eqref{eqn:L_superoperator}, and throughout the remainder of this paper, we use the notational shorthand \begin{equation}
    A^{\oplusbar 2} \equiv A \otimes I + I \otimes \overline{A}
\end{equation}
to denote the conjugated Kronecker sum and 
\begin{equation}
    A^{\otimesbar 2} \equiv A \otimes \overline{A}
\end{equation} 
to denote the conjugated Kronecker product. \\

It is worth remarking that the effective Hamiltonian superoperator $\mathcal{H}$ does not generate a unitary evolution of the density matrix, though it can be written in a form that evolves $\rho$ in a manner analogous to that of unitary evolution under the effective Hamiltonian $V_H$:

\begin{equation}
e^{t\mathcal{H}}\vec{\rho}\quad \Leftrightarrow\quad (e^{-it V_H / \hbar}) \rho (e^{-it V_H / \hbar})^{\dagger}
\label{eqn:H_superop_dynamics}
\end{equation}

The dynamics of \eqref{eqn:H_superop_dynamics} are non-unitary because the damping terms $\frac{-i\hbar}{2} \sum_n \gamma_n L_n^\dagger L_n$ in \eqref{eqn:V_H} are skew-Hermitian with non-positive purely imaginary eigenvalues. As a result, the trace and purity of $\rho$ are not preserved with time. However, in the no-damping limit where $\gamma_n~\rightarrow~0$, these terms vanish and $\mathcal{H}$ generates unitary dynamics.

Evolving the system under $\mathcal{H}$ alone is insufficient to account for the direct action of the environment on the system. To completely account for this action, we must also incorporate the Lindblad superoperator $\mathcal{L}$, given in \eqref{eqn:L_superoperator}. If we combine the action of $\mathcal{H}$ and $\mathcal{L}$ on a vectorized density matrix $\vec{\rho}$, we obtain a linear superoperator equivalent to the right hand side of the Lindblad equation \eqref{eqn:lindblad}, which is completely positive and trace-preserving:

\begin{equation}
    \dfrac{d\vec{\rho}}{dt}(t) = (\mathcal{H} + \mathcal{L})\vec{\rho}(t).
    \label{eqn:super_lindblad}
\end{equation}

We can verify that the equivalence of \eqref{eqn:super_lindblad} to \eqref{eqn:lindblad} holds by expanding $\mathcal{H}$ and $\mathcal{L}$ to obtain \eqref{eqn:super_lindblad_rhs_expansion}, and checking the operator-to-superoperator correspondence in \eqref{eqn:superops}:

\begin{widetext}
\begin{equation}
 \mathcal{H} + \mathcal{L} = \left( \frac{-i}{\hbar}(H \otimes I - I \otimes \overline{H}) - \frac{1}{2}\sum_n \gamma_n(L_n^\dagger L_n \otimes I + I \otimes \overline{L_n}^\dagger\overline{L_n}) \right) + \left( \sum_n \gamma_n (L_n \otimes \overline{L_n})\right).
\label{eqn:super_lindblad_rhs_expansion}
\end{equation}
\end{widetext}

Since ($\mathcal{H}$ + $\mathcal{L}$) acts linearly on $\vec{\rho}$, the time-evolution of the system is given by
\begin{equation}
    \vec{\rho}(t) = e^{t(\mathcal{H} + \mathcal{L})}\vec{\rho}(0).
\label{eqn:super_lindblad_exponential}
\end{equation}

We know that $e^{t(\mathcal{H} + \mathcal{L})}$ must be equal to a sum of time-dependent Kraus operators $K_i(t)$, which means that such a representation satisfies the equation
\begin{equation}
\sum_{i} K_i(t)^{\otimesbar 2} = e^{t(\mathcal{H} + \mathcal{L})}.
\label{eqn:kraus_superop_correspondence}
\end{equation}

Finding a set of $K_i(t)$ in a closed matrix form that satisfies this for all $t \ge 0$ is a nontrivial task, as it amounts to decomposing an operator exponential into a sum of terms that factor as a conjugated Kronecker product (i.e. $K_i(t)^{\otimesbar 2}$). In the limit of an infinitesimally small time-step $\delta t$, however, we can recover a set of first order generators for $N_L + 1$ Kraus operators corresponding to a set of $N_L$ Lindblad operators. These are of the form:

\begin{align}
    K_{0}(\delta t) &= I + \frac{-i(\delta t)}{\hbar}V_H
    \label{eqn:K_H_numerical} \\
    K_n(\delta t) &= \sqrt{\delta t}V_n\qquad (\text{for } n = 1, ..., N_L).\label{eqn:K_n_numerical}
\end{align}

The generator \eqref{eqn:K_H_numerical} can be integrated in closed form to obtain the evolution of the system under the effective Hamiltonian, $e^{t\mathcal{H}}$. Specifically, it can be shown that
\begin{equation}
K_0(t) = e^{-it V_H/\hbar},
\label{eqn:trivial_kraus_operator}
\end{equation}
which, when substituted into \eqref{eqn:krauss}, is equivalent to \eqref{eqn:H_superop_dynamics}. Unlike $K_0(\delta t)$, the generators of the form \eqref{eqn:K_n_numerical} do not correspond to closed-form Kraus operators, and thus must be integrated numerically, either through Trotter-based techniques on a quantum computer, or through standard numerical methods on a classical computer. In either case, we consider these methods to be less desirable for NISQ computing applications. This motivates our search for classes of systems that admit Kraus representations that satisfy \eqref{eqn:kraus_superop_correspondence} and can be represented as a closed-form function of time, even if such representations require more Kraus operators than those that can be computed numerically.

\subsubsection{The Environmental Interaction Picture}
\label{sec:environmental_interaction_picture}

In this paper, we take a different approach to solving for a Kraus representation of the system. Instead of using the minimal set of $N_L+1$ Kraus operators generated by \eqref{eqn:K_H_numerical} and \eqref{eqn:K_n_numerical}, we solve for an infinite series of Kraus operators which can be derived by applying the Zassenhaus product formula to the right-hand side of \eqref{eqn:kraus_superop_correspondence}. The Zassenhaus product formula expands $e^{t(\mathcal{H} + \mathcal{L})}$ as the product of exponentials of commutators of $\mathcal{H}$ and $\mathcal{L}$ as follows:

\begin{equation}
\begin{aligned}
    e^{t(\mathcal{H} + \mathcal{L})} &= e^{t\mathcal{H}}e^{t\mathcal{L}}e^{-\frac{t^2}{2}[\mathcal{H},\mathcal{L}])}\\
    &\qquad \times e^{\frac{t^3}{6}([\mathcal{H},[\mathcal{H},\mathcal{L}]] + 2[\mathcal{L},[\mathcal{H},\mathcal{L}]])}e^{-\frac{t^4}{24}(...)}... \\
    &= e^{t\mathcal{H}} \prod_{m=1}^\infty \exp\left(t^m \mathcal{C}_m\right).
    \label{eqn:zassenhaus}
\end{aligned}
\end{equation}
For $m>1$, $\mathcal{C}_m$ denotes the degree $m$ homogeneous Lie polynomial exponents in the Zassenhaus expansion, whereas for $m=1$ we fix $\mathcal{C}_1 := \mathcal{L}$. These exponents can be obtained by recursively applying the Baker-Campbell-Hausdorff formula, or through one of the methods outlined in \cite{dupays_closed_2023} or \cite{casas_efficient_2012}. In \tabref{\ref{tab:zassenhaus_terms}}, we list the commutator exponents $\mathcal{C}_m$ up to $m=5$.

\begin{table}[]
    \centering
    \begin{tabular}{ c  c}
    \hline\hline
     $~m~$ &  $\mathcal{C}_m$ \\
     \hline \\[-3mm]
     $1$ & $\mathcal{L}$ \\[2mm]
     $2$ & $-\frac{1}{2}[H,L]$ \\[2mm]
     $3$ & $\frac{1}{6}([\mathcal{H},[\mathcal{H},\mathcal{L}]] + 2[\mathcal{L},[\mathcal{H},\mathcal{L}]])$ \\[2mm]
     $4$ & $-\frac{1}{24}([\mathcal{H},[\mathcal{H},[\mathcal{H},\mathcal{L}]]] + 3[\mathcal{L},[\mathcal{H},[\mathcal{H},\mathcal{L}]]] + 3[\mathcal{L},[\mathcal{L}, [\mathcal{H},\mathcal{L}]]])$ \\[2mm]
     $5$ & $\begin{aligned}
            {\tfrac{1}{120}}(&[\mathcal{H},[\mathcal{H},[\mathcal{H},[\mathcal{H},\mathcal{L}]]]]+ 4[\mathcal{L},[\mathcal{H},[\mathcal{H},[\mathcal{H},\mathcal{L}]]]])~+\\
            &6[\mathcal{L},[\mathcal{L},[\mathcal{H},[\mathcal{H},\mathcal{L}]]]] + 4[\mathcal{L},[\mathcal{L},[\mathcal{L},[\mathcal{H},\mathcal{L}]]]])~+\\
     &6[[\mathcal{H},\mathcal{L}],[\mathcal{H},[\mathcal{H},\mathcal{L}]]] + 12[[\mathcal{H},\mathcal{L}],[\mathcal{L},[\mathcal{H},\mathcal{L}]]])
     \end{aligned}$ \\[6mm]
     \hline\hline
\end{tabular}
    \caption{Table of the Zassenhaus formula commutator exponents $\mathcal{C}_m$ in \eqref{eqn:zassenhaus} up to $m=5$.}
    \label{tab:zassenhaus_terms}
\end{table}

Equation \eqref{eqn:zassenhaus} re-frames the right hand side of \eqref{eqn:kraus_superop_correspondence} as the evolution of a product of Zassenhaus commutator exponentials $(\prod_{m=1}^{\infty} \exp(t^m \mathcal{C}_m))$ under the effective Hamiltonian superoperator $\mathcal{H}$. This motivates the introduction of the environment action superoperator
\begin{equation}
\mathcal{A}_{\text{env}}(t) = \prod_{m=1}^{\infty} e^{t^m\mathcal{C}_m},
\label{eqn:A_superoperator}
\end{equation}
which incorporates the cumulative effect of the Lindblad superoperator $\mathcal{L}$ acting on the system over a time period $t > 0$. We can interpret $\mathcal{A}_{\text{env}}(t)$ as a superoperator that evolves the system in the ``environmental interaction picture", which is related to the Schr\"odinger picture by the non-unitary (but nonetheless invertible) transformation $e^{t\mathcal{H}}$. It immediately follows that the time-evolution of the system can be written in the form
\begin{equation}
    \vec{\rho}(t) = e^{t\mathcal{H}}\mathcal{A}_{\text{env}}(t)\vec{\rho}(0).
    \label{eqn:interaction_density_evolution}
\end{equation}

\begin{figure}
    \centering
    \includegraphics[width=\linewidth]{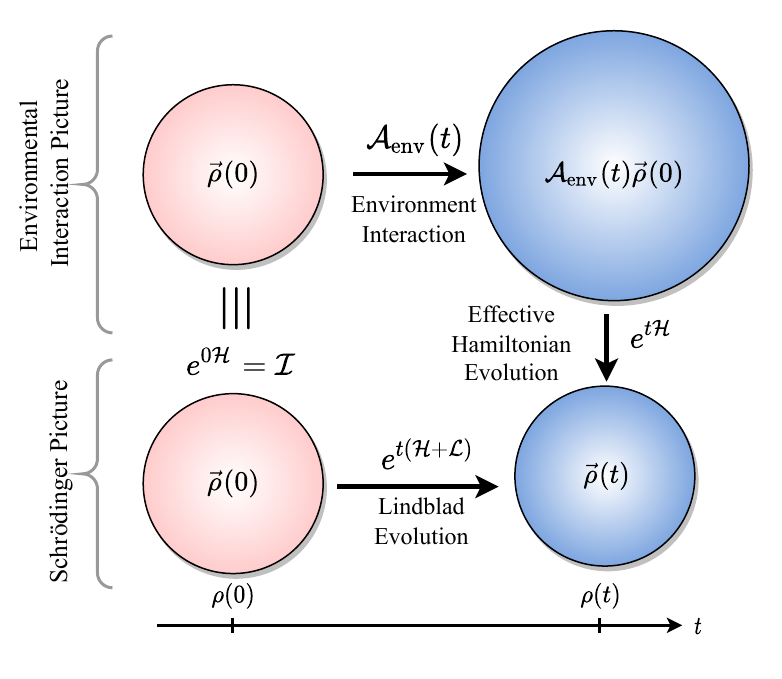}
    \caption{Illustration of the relationship between the evolution of the system in the Schr\"odinger picture and the environmental interaction picture. 
    In the environmental interaction picture, the system is evolved with the superoperator $\mathcal{A}_{\text{env}}(t)$, which is not trace-preserving. This evolved state can be transformed back to the Schrodinger picture through evolution under the effective Hamiltonian, which takes the form \eqref{eqn:H_superop_dynamics}. The state $\mathcal{A}_{\text{env}}(t)\vec{\rho}(0)$ is shown larger than the others to emphasize that the trace of the density operator is not preserved in the environmental interaction picture. The transformation $e^{t\mathcal{H}}$ renormalizes the trace and returns the system to the Schr\"odinger picture.}
    \label{fig:evolution_diagram}
\end{figure}

If both $e^{t\mathcal{H}}$ and $\mathcal{A}_{\text{env}}(t)$ can be realized efficiently on a quantum computer, then the time evolution of the entire system can be computed efficiently by first evolving $\rho(0)$ in the environmental interaction picture using $\mathcal{A}_{\text{env}}(t)$ and then transforming back into the Schr\"odinger picture as suggested in \figref{\ref{fig:evolution_diagram}}. However, computing $\mathcal{A}_{\text{env}}(t)$ for arbitrary systems may be quite difficult, since it requires computing the product of all Zassenhaus exponentials of the form $\exp(t^m\mathcal{C}_m)$. In the next section we will show that when a system satisfies certain superoperator commutation relations, these commutator exponentials reduce to a tractable form, which allows for the derivation of a closed form series expansion of $\mathcal{A}_{\text{env}}(t)$.

\subsection{Derivation of Kraus Operators}
\label{sec:derivation_of_kraus_operators}

In \secref{\ref{sec:introduction}}, we claimed to have found a class of systems which admit a Kraus operator series representation, and we characterized them as satisfying three general conditions (conditions 1-3). For now we will defer the discussion of conditions 1 and 2 to a later section and focus on condition 3, which requires that a system's $\mathcal{H}$ and $\mathcal{L}$ superoperators satisfy the commutation relation
\begin{equation}
    [ \mathcal{H}, \mathcal{L} ] = \alpha \mathcal{L} + c
    \label{eqn:property_3_commutator}
\end{equation}
for $\alpha, c \in \mathbb{R}_{\ge 0}$. We will proceed by first deriving closed form Kraus operators for systems satisfying condition 3, and then show how they can be simulated with quantum circuits. To facilitate the derivation, we will consider the following two cases separately:\\
\begin{align}
    \text{Case (I):}\quad & [\mathcal{H},\mathcal{L}] = c\ \qquad (\text{for } c \ge 0) \label{eqn:case_1}\\[4mm]
    \text{Case (II):}\quad &[\mathcal{H}, \mathcal{L}] = \alpha \mathcal{L} + c \quad (\text{for } \alpha > 0, c \ge 0) \label{eqn:case_2}
\end{align}

\subsubsection{Case (I) Kraus Representation}
\label{sec:case_1_kraus_representation}

For case (I) systems, we apply the commutation relation \eqref{eqn:case_1} to the Zassenhaus formula \eqref{eqn:zassenhaus} and observe that all commutator exponents $\mathcal{C}_{m}$ vanish except for $\mathcal{C}_1 = \mathcal{L}$ and $\mathcal{C}_2 = -\frac{c}{2}$.
\begin{equation}
    e^{t(\mathcal{H} + \mathcal{L})} = e^{t\mathcal{H}}e^{t\mathcal{L}}e^{-\frac{ct^2}{2}}
    \label{case_1_zassenhaus}
\end{equation}
Substituting the non-vanishing $\mathcal{C}_m$ terms into \eqref{eqn:A_superoperator}, we see that the environment action superoperator is
\begin{equation}
    \mathcal{A}_{\text{env}}(t) =  e^{-ct^2/2}e^{t\mathcal{L}}.
    \label{eqn:case_1_A_env}
\end{equation}
Expanding the $e^{t\mathcal{L}}$ in $\mathcal{A}_{\text{env}}(t)$ as a Taylor series, we obtain
\begin{equation}
    \mathcal{A}_{\text{env}}(t) = \sum_{m=0}^{\infty} \frac{t^m}{m!}\mathcal{A}_{\text{env}}^{(m)}(t),
    \label{eqn:A_env_series}
\end{equation}
where each term $\mathcal{A}_{\text{env}}^{(m)}$ can be further expanded as a sum over all product sequences of $m$ Lindblad operators:
\begin{align}
    \mathcal{A}_{\text{env}}^{(m)}(t) &= e^{-\frac{ct^2}{2}}\mathcal{L}^m \label{eqn:case_1_exp_L}\\
    &= e^{-\frac{ct^2}{2}}\left( \sum_{n=1}^N V_n^{\otimesbar 2} \right)^m \label{eqn:case_1_exp_L_series}\\
    &= e^{-\frac{ct^2}{2} }\sum_{\vec{k} \in \{1, 2, ..., N_L\}^{m}} \left( \prod_{j=1}^m \left[\sqrt{\gamma_{\vec{k}_j}} L_{\vec{k}_j} \right] \right)^{\otimesbar 2} 
    \label{eqn:case_1_exp_L_expanded}
\end{align}
Each of the $(N_L)^m$ summation terms in \eqref{eqn:case_1_exp_L_expanded} is indexed by a vector $\vec{k}\in \{1, 2, ..., N_L\}^{m}$. This vector indicates which $m$ (out of the total $N_L$) Lindblad operators are applied, and in which order.

By combining equations \eqref{eqn:interaction_density_evolution}, \eqref{eqn:A_env_series}, and \eqref{eqn:case_1_exp_L_expanded}, we can write the time evolution of the system as a Kraus series that satisfies \eqref{eqn:kraus_superop_correspondence}. This series takes the superoperator form
\begin{equation}
    \vec{\rho}(t) = \sum_{m=0}^{\infty} \sum_{\vec{k} \in \{1,2,...,N_L\}} K_{m,\vec{k}}(t)^{\otimesbar 2}\vec{\rho}(0),
    \label{eqn:lindblad_superoperator_kraus_series}
\end{equation}
which is equivalent to
\begin{equation}
    \rho(t) = \sum_{m=0}^{\infty} \sum_{\vec{k} \in \{1,2,...,N_L\}} K_{m,\vec{k}}(t)\rho(0) K_{m,\vec{k}}^{\dagger}(t).
    \label{eqn:lindblad_kraus_series}
\end{equation}
The Kraus operators in the series are indexed by both a perturbative order $m$ and a vector $\vec{k}$ corresponding to a sequence of Lindblad operators in the expansion of $\mathcal{A}_{\text{env}}^{(m)}$. The Kraus operators for the case (I) system are
\begin{equation}
    K_{m, \vec{k}}^{(\text{I})}(t) = e^{\frac{-it}{\hbar}V_H} \sqrt{\frac{t^m e^{-\frac{ct^2}{2}}}{m!}} \prod_{j=1}^m \left(\sqrt{\gamma_{\vec{k}_j}} L_{\vec{k}_j}\right).
    \label{eqn:case_1_kraus_operator}
\end{equation}
In the case of a single Lindblad operator $L_1$, the index vector $\vec{k}$ can be removed and the Kraus operators reduce to the much simpler form
\begin{equation}
    K_m^{(\text{I})}(t) = e^{\frac{-it}{\hbar}V_H} \sqrt{\frac{(\gamma_1 t)^me^{-\frac{ct^2}{2}}}{m!}}  L_1^m
    \label{eqn:case_1_single_kraus_operator}
\end{equation}
for $m = 0, 1, 2, ...$ etc.

\subsubsection{Case (II) Kraus Representation}
\label{sec:case_2_kraus_representation}

For case (II) systems, we apply the commutation relation \eqref{eqn:case_2} and observe that $\mathcal{C}_m \propto t^m \mathcal{L}$ for all commutator exponent terms $m > 2$ in the Zassenhaus expansion. In particular, it can be shown that \eqref{eqn:zassenhaus} collapses to
\begin{align}
e^{t(\mathcal{H} + \mathcal{L})} &= e^{t\mathcal{H}}e^{(t + \alpha g(t, \alpha))\mathcal{L}}e^{cg(t,\alpha)},
\label{eqn:case_1_exponential}
\end{align}
so we obtain
\begin{equation}
    \mathcal{A}_{\text{env}}(t) = e^{(t + \alpha g(t, \alpha))\mathcal{L}}e^{cg(t,\alpha)},
\end{equation}
\noindent where $g(t, \alpha)$ (derived in \cite{dupays_closed_2023}) has the closed form 
\begin{equation}
    g(t, \alpha) = -\frac{e^{-\alpha t} + \alpha t - 1}{\alpha^2}.
\end{equation}

Following the same procedure as in case (I), we expand $\mathcal{A}_{\text{env}}(t)$ as the series
\begin{equation}
        \mathcal{A}_{\text{env}}(t) = \sum_{m=0}^{\infty} \frac{(t + \alpha g(t,\alpha))^m}{m!}\mathcal{A}_{\text{env}}^{(m)}(t),
    \label{eqn:A_env_series_2}
\end{equation}
where
\begin{equation}
    \begin{aligned}
    \mathcal{A}_{\text{env}}^{(m)}(t) = e^{\frac{cg(t,\alpha)}{2} }\sum_{\vec{k} \in \{1, 2, ..., N_L\}^{m}} \left( \prod_{j=1}^m \left[\sqrt{\gamma_{\vec{k}_j}} L_{\vec{k}_j} \right] \right)^{\otimesbar 2}.
    \end{aligned}
    \label{eqn:case_2_exp_L_expanded}
\end{equation}

After simplifying each term, we obtain a Kraus series of the form \eqref{eqn:lindblad_kraus_series} where the Kraus operators $K_{m, \vec{k}}^{(\text{II})}(t)$ are given by
\begin{equation}
    K_{m, \vec{k}}^{(\text{II})}(t) = e^{\frac{-it}{\hbar}V_H} \sqrt{\frac{(1 - e^{-\alpha t})^m e^{cg(t, \alpha)}}{\alpha^m ~m!}} \prod_{j=1}^m \left(\sqrt{\gamma_{\vec{k}_j}} L_{\vec{k}_j}\right).
    \label{eqn:case_2_kraus_operator}
\end{equation}
In the case of a single Lindblad operator $L_1$, the Kraus operators reduce to
\begin{equation}
\begin{aligned}
    K_m^{(\text{II})}(t) &= e^{\frac{-it}{\hbar}V_H} \sqrt{\frac{\gamma_1^m(1 - e^{-\alpha t})^m e^{cg(t, \alpha)}}{\alpha^m ~m!}}  L_1^m
    \end{aligned}
    \label{eqn:case_2_single_kraus_operator}
\end{equation}
for $m = 0, 1, 2, ...$ etc.

\begin{figure}
    \centering
    \includegraphics[width=\linewidth]{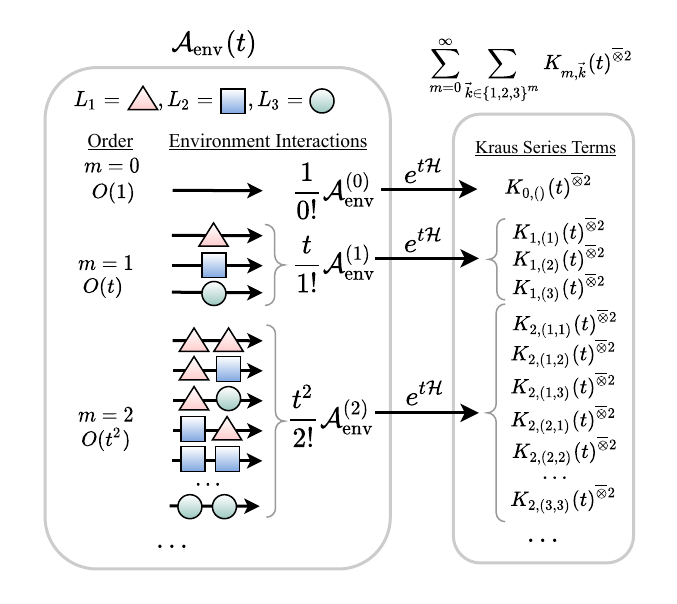}
    \caption{Illustration of the mapping between the environment action superoperator $\mathcal{A}_{\text{env}}(t)$ and the Kraus operator series of the form \eqref{eqn:lindblad_kraus_series} for systems satisfying the superoperator commutation relation \eqref{eqn:property_3_commutator}. In the environmental interaction picture, each sequence of $m$ Lindblad operators (which models a specific mode of environmental interaction) corresponds to an order $m$ term in a Kraus series that evolves the system in the Schr\"odinger picture.}
    \label{fig:kraus_mapping}
\end{figure}

\subsection{Interpreting and Simplifying the Kraus Series}
\label{sec:simplifying_the_kraus_series}

We have derived a Kraus series representation of the system where each Kraus operator $K_{m,\vec{k}}(t)$ can be interpreted as the evolution of an ordered sequence of Lindblad operators (i.e. $\prod_{i} L_{\vec{k}_i} = L_{\vec{k}_1}L_{\vec{k}_2}...L_{\vec{k}_m}$) under the system's effective Hamiltonian. For the $m=0$ Kraus operator, this sequence is empty and thus corresponds to evolution under the effective Hamiltonian only, as given by \eqref{eqn:trivial_kraus_operator}. The remaining Kraus operators contain nonempty sequences of $m > 0$ Lindblad operators. Each sequence can be interpreted as a mode of environmental interaction in which a number of physical processes occur in a specific order, such as particle exchange with the environment, environment-mediated transitions in state, or the creation and subsequent annihilation of virtual particles within the system. Through this lens, the terms of the Kraus series can be viewed in a manner analogous to the particle-path interpretation that arises from the time-perturbative treatment of a scattering process \cite{peskin_introduction_1995}. However, accounting for all order $m$ modes of interaction in a system with many Lindblad operators requires a number of Kraus operators that grows exponentially with $m$ unless one can simplify the Kraus series using commutator and product identities associated with the Lindblad operators.

The most straightforward simplification case to consider is when the Lindblad operators commute (i.e. $[L_i, L_j] = 0$, or more generally $[L_i^{\otimesbar 2}, L_j^{\otimesbar 2}] = 0$). This allows for the set of index vectors $\vec{k}$ to be grouped into equivalent sets according the number of times each Lindblad operator appears in the sequence $\prod_{i} L_{\vec{k}_i}$. Letting $\vec{\ell}$ denote a representative from each of these sets of $\vec{k}$ vectors for every order $m$, the Kraus series can be reduced to
\begin{equation}
    \vec{\rho}(t) = \sum_{m=0}^{\infty} \sum_{\vec{\ell}} c_{m,\vec\ell} K_{m,\vec{l}}^{\otimesbar 2} \vec{\rho},
\end{equation}
where $c_{m,\ell}$ is a multinomial coefficient
\begin{equation}
    c_{m,\ell} = \binom{m}{n_1(\vec\ell), n_2(\vec\ell), ..., n_{N_L}(\vec\ell)}
\end{equation}
which counts the number of ways $n_1(\vec\ell), n_2(\vec\ell), ..., n_{N_L}(\vec\ell)$ instances of the respective operators $L_1, L_2, ..., L_{N_L}$ can be ordered to create an equivalent product sequence of length $m = n_1(\vec\ell) + n_2(\vec\ell) + ... + n_{N_L}(\vec\ell)$. This reduces the number of Kraus series terms of each order $m$ by a factor roughly proportional to $(N_L)^m/m!$ for $m > N_L$.

The Kraus series can also be reduced by incorporating product identities of the form $L_iL_j = L_k$. This most commonly occurs when a system's Lindblad operators form a multiplicative group up to multiplication by a constant. When the Lindblad operators are either nilpotent or are endowed with a finite group structure, it is possible to derive very useful power formulas of the form $L_i^{n} = \theta I$ (or more generally $(L_i^{\otimesbar 2})^n = |\theta|^2I^{\otimesbar 2}$), where $\theta$ is some constant and $I$ is the identity operator. These can be used to reduce the Kraus series to a finite number of terms up to a maximum order $n$. For example, with case (I) systems containing a single Lindblad operator $L_1$ satisfying $L_1^{n} = \theta I$, we can reduce the the infinite series of Kraus operators of the form \eqref{eqn:case_1_single_kraus_operator} down to the finite sum of Kraus operators
\begin{equation}
    K_{m}^{(I)}(t) = e^{\frac{-it}{\hbar}V_H}\sqrt{F_{n,m}^{\theta}(\gamma_1 t)e^{-\frac{ct^2}{2}}} L_1^m
\end{equation}
for $m = 0, 1, 2, ..., n-1$. A similar form can be found for case (II) systems. Above, $F_{n,m}^{\theta}(x)$ denotes the generalized hyperbolic function 
\begin{equation}
    F_{n,m}^{\theta}(x) = \sum_{k=0}^{\infty} \frac{\theta^k}{(nk + m)!}x^{nk + m},
\end{equation}
which can be computed in closed form (following the method in \cite{muldoon_generalized_2005}) using the formula
\begin{equation}
    F_{n,m}^{\theta}(x) = \frac{1}{n}\theta^{-\frac{m}{n}} \sum_{k=0}^{n-1} \omega_n^{-mk} \exp(\omega_n^{k} \theta^{\frac{1}{n}} x),
\end{equation}
where $\omega_{n} = e^{i2\pi/n}$ is an $n$-th primitive root of unity.

\subsection{Representing Kraus Operators as Quantum Circuits}
\label{sec:representing_kraus_operators_as_quantum_circuits}

So far, we have shown that open quantum systems satisfying the commutation relation \eqref{eqn:case_1} or \eqref{eqn:case_2} admit a Kraus series representation, yet it remains to be discussed how the representations can be realized as quantum circuits. Since the Kraus operators $K_i(t)$ for a system are generally non-unitary, they must first be expanded to unitary operators to be implemented on a quantum computer. These expanded unitary operators can be viewed as acting on a representation of the system plus an ancillary environmental subsystem. This expansion is usually achieved though either a Stinespring dilation or a Sz.-Nagy dilation of the Kraus operators.

\subsubsection{Stinespring Dilations}
\label{sec:stinespring_dilations}

The most common of unitary dilation methods is the Stinespring dilation \cite{stinespring_positive_1955}, which uses the total unitary evolution of a set of system qubits which become coupled with a set of environmental qubits. It represents the total system as the tensor product of the system and environmental degrees of freedom, and requires finding a unitary $U_{\text{tot}}(t)$ that satisfies
\begin{equation}
    \begin{aligned}
    &\Tr_{\text{env}}(U_{\text{tot}}(t)(\rho_{\text{sys}} \otimes \rho_{\text{env}})U_{\text{tot}}(t)^\dagger) \\
    &\qquad\qquad\qquad = \sum_{i} K_{i}(t)\rho_{\text{sys}} K_i(t)^\dagger,
    \end{aligned}
    \label{eqn:kraus_stinespring_dilation}
\end{equation}
where $\rho_{\text{env}} = \ket{\phi}\bra{\phi}_{\text{env}}$ is a pure state of rank one and $\rho_{\text{sys}}$ is an arbitrary initial state. 

\begin{figure}[h!]
    \centering
    \includegraphics[width=0.85\linewidth]{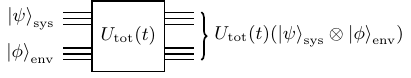}
    \caption{Quantum circuit diagram of a Stinespring dilation circuit simulating a system initialized in the pure state $\ket{\psi}_{\text{sys}}$.}
    \label{fig:stinespring_dilation_circuit}
\end{figure}

If the initial state $\rho_{\text{sys}}$ is pure (i.e. $\Tr(\rho_{\text{sys}}^2) = 1$), we can decompose $\rho_{\text{sys}} = \ket{\psi}\bra{\psi}_{\text{sys}}$ and simulate the evolution of the entire system by evolving $\ket{\psi}_{\text{sys}}$ only. This is done by applying $U_{\text{tot}}(t)$ to the system state $\ket{\psi}_{\text{sys}}$ and environment state $\ket{\phi}_{\text{env}}$ as shown in \figref{\ref{fig:stinespring_dilation_circuit}}, after which quantities such as state occupation probabilities or the expectation values of observables can be estimated by measuring the system qubits only. This effectively traces out the environment subsystem, in accordance with \eqref{eqn:kraus_stinespring_dilation}.

In the case that $\rho_{\text{sys}}$ is not pure, we can still simulate its evolution by diagonalizing it in the form
\begin{equation}
    \rho_{\text{sys}} = \sum_{j} p_j \ket{\psi_j}\bra{\psi_j}_{\text{\text{sys}}}
\end{equation}
and then preparing a purification of $\rho_{\text{sys}}$ via a set of ancillary prep qubits indexed by an orthonormal basis $\ket{e_j}$:
\begin{equation}
    \rho_{\text{sys}} \sim \sum_{j} \sqrt{p_j} (\ket{\psi_j}_{\text{sys}} \otimes \ket{e_j}_{\text{prep}}).
    \label{eqn:purified_state_prep}
\end{equation}
This prepares a superposition of $\ket{\psi_j}_{\text{sys}}$ states that can used in place of $\ket{\psi}_{sys}$ in circuits such as \figref{\ref{fig:stinespring_dilation_circuit}}.

\subsubsection{Sz.-Nagy Dilations}
\label{sec:sz_nagy_dilations}

Unlike a Stinespring dilation, a Sz.-Nagy dilation represents the total system as the direct sum of the system and the environmental degrees of freedom for each individual Kraus operator \cite{schaffer_unitary_1955}. The unitary Sz.-Nagy dilation of a Kraus operator $K_i(t)$ is given by
\begin{equation}
    U_{K_i}(t) = \begin{pmatrix}
        K_i(t)     & D_{K_i^\dagger}(t) \\
        D_{K_i}(t) & -K_i(t)^\dagger \\
    \end{pmatrix}
    \label{eqn:sznagy_dilation}
\end{equation}
where $D_{K_i}(t)$ is the defect operator of $K_i(t)$, defined as
\begin{equation}
    D_{K_i}(t) =  \sqrt{I - K_i(t)^\dagger K_i(t)}.
\end{equation}

\begin{figure}[h]
    \centering
    \includegraphics[width=0.85\linewidth]{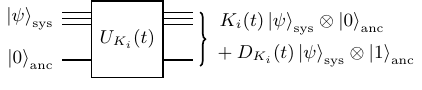}
    \caption{Quantum circuit diagram of a Sz.-Nagy dilation quantum circuit for the Kraus operator $K_i(t)$.}
    \label{fig:sznagy_dilation_circuit}
\end{figure}

Representing $U_{K_i}$ on a quantum computer requires only a single ancillary qubit. This qubit distinguishes the system degrees of freedom from the environmental degrees of freedom by its measured value ($\ket{0}$ for the system, and $\ket{1}$ for the environment). To compute expectation values of system observables $\hat{O}_{\text{sys}}$ with respect to the system's final state $\rho(t)$, one must construct a quantum circuit representation of $U_{K_i}(t)$ for each Kraus operator $K_i(t)$ as shown in \figref{\ref{fig:sznagy_dilation_circuit}} and then compute the corresponding expectation values for each circuit while filtering out any $\ket{1}$ measurements in the ancillary qubits. This is done by defining the filtered observable 
\begin{equation}
    \hat{O}' = (\hat{O}_{\text{sys}} \otimes \ket{0}\bra{0}_{\text{anc}})
\end{equation}
and measuring the expectation value $\langle \hat{O}'\rangle_{K_i(t)}$ on both the system and Sz.-Nagy ancillary qubits for each Kraus circuit $K_i$. The resulting expectation values for each Kraus operator circuit are then added together (on a classical computer) to recover the system expectation value:
\begin{equation}
\langle \hat{O}_{\text{sys}} \rangle_{\rho(t)} = \sum_{i} \langle \hat{O}'\rangle_{{K_i}(t)}.
\end{equation}

If any ancilla qubits are used aside from those needed to realize the Sz.-Nagy dilations contained in a circuit (for instance, qubits used to prepare the purification of the mixed state in \eqref{eqn:purified_state_prep}) these qubits are ignored and not measured in the evaluation of $\langle \hat{O}' \rangle_{K_i(t)}$.

A key advantage of using Sz.-Nagy dilations is the lower representation complexity. Indeed, it is known that if the representation of a given Kraus operator is minimal, both the minimality and uniqueness (up to a unitary transform) of the representation $U_{K_i}(t)$ is guaranteed, per the Sz.-Nagy dilation theorem \cite{schaffer_unitary_1955}. As a result Sz.-Nagy dilations typically require fewer two-qubit entangling gates to implement in comparison to Stinespring dilations, as the dimension of the dilated unitaries in the form of \eqref{eqn:sznagy_dilation} tend to be orders of magnitude less than those of a $U_{tot}(t)$ unitary that explicitly models environmental interactions. The disadvantage of using Sz.-Nagy dilations, however, is that they split the system dynamics over many quantum circuits, requiring one circuit per Kraus operator, which may be undesirable for systems with many Kraus operators. On NISQ devices, we argue that Sz.-Nagy dilations are preferable to Stinespring dilations, due to their lower gate complexity, which mitigates the effect of simulation noise, thereby allowing for more complex systems to be simulated.

\subsubsection{Representation of case (I) and case (II) Kraus operators.}
\label{sec:representation_of_kraus_operators}

Now we will focus on the problem of representing the Kraus operators for the special cases (I) and (II), for which we derived the closed forms of \eqref{eqn:case_1_kraus_operator} and \eqref{eqn:case_2_kraus_operator} respectively. A key assumption we have made for these systems (as stated in condition 2 in the introduction) is that the effective Hamiltonian dynamics (i.e. $\exp(t\mathcal{H})$) can be efficiently represented as a quantum circuit of $t$-independent gate complexity. Per \eqref{eqn:H_superop_dynamics}, this means that it is an equivalent condition for $e^{\frac{-it}{\hbar}V_H}$ to be efficiently realizable with $t$-independent gate complexity. One such realization is given by the Sz.-Nagy dilation

\begin{equation}
    U_\mathcal{H}(t) = \begin{pmatrix}
        e^{\frac{-it}{\hbar}V_H} &D_{\exp(\frac{-it}{\hbar} V_H)^{\dagger}} \\[2mm]
        D_{\exp(\frac{-it}{\hbar} V_H)} & -(e^{\frac{-it}{\hbar}V_H})^{\dagger}.
    \end{pmatrix}
    \label{eqn:U_H_sznagy_dilation}
\end{equation}

\begin{figure}[h!]
    \centering
    \includegraphics[width=\linewidth]{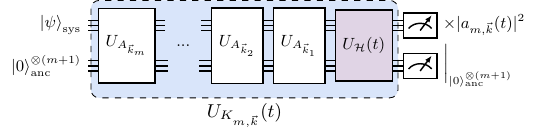}
    \caption{Quantum circuit diagram for the general multi-step implementation of the case (I) and case (II) Kraus operators $K_{m,\vec{k}}(t)$. First, a sequence of scaled Lindblad operators $A_{i}$ are applied according to the index ordering $i=\vec{k}_m,\vec{k}_{m-1},...,\vec{k}_1$. Finally, $U_{\mathcal{H}}$ is applied and system and ancilla qubits are measured. When the ancillas all read $\ket{0}$ the measured probabilities (or expectation values) are re-scaled by $|a_{m,\vec{k}}(t)|^2$. If any of the ancillas are not $\ket{0}$, then the probability of that state is discarded.}
    \label{fig:kraus_operator_circuit}
\end{figure}

Now that we have addressed how $U_{\mathcal{H}}(t)$ can be represented, we can realize a full Kraus operator $K_{m, \vec{k}}(t)$ of the form \eqref{eqn:case_1_kraus_operator} or \eqref{eqn:case_2_kraus_operator} as a quantum circuit by first applying a sequence of Sz.-Nagy dilations $U_{A_{\vec{k}_i}}$ (each of which apply a re-scaled form of the the $\vec{k}$-indexed Lindblad operators $L_{\vec{k}_i}$) followed by $U_{\mathcal{H}}$. The general layout of this circuit is shown in \figref{\ref{fig:kraus_operator_circuit}}.

The re-scaling of the Lindblad operators is necessary, since it is not generally guaranteed to be the case that $\norm{L_i} \le 1$. Thus we define the re-scaled Lindblad operators
\begin{equation}
    A_{i} = \frac{L_{i}}{\norm{L_{i}}}
    \label{eqn:rescaled_lindblad_ops}
\end{equation}
and use them to construct the Sz.-Nagy dilations $U_{A_i}$, as shown in \figref{\ref{fig:kraus_operator_circuit}}. This can be done with standard unitary decomposition methods. While scaling is necessary to make the $L_i$ operators realizable as quantum circuits, it requires that any measured observables under the final wave function must be subsequently unscaled by a squared factor ($|a_{m,\vec{k}}(t)|^2$) to obtain the correct value. The factor $a_{m,\vec{k}}(t)$ is time-dependent, because we also use it to incorporate all of the remaining time-dependent scalar values contained in the closed-form Kraus operators. Specifically, for the general case (I) Kraus operators of the form \eqref{eqn:case_1_kraus_operator}, we use
\begin{equation}
    a_{m,\vec{k}}^{(\text{I})} = \sqrt{\frac{t^m e^{-\frac{ct^2}{2}}}{m!}} \prod_{j=1}^m \sqrt{\gamma_{\vec{k}_j}} \norm{L_{\vec{k}_j}},
\end{equation}
and likewise for the case (II) Kraus operators of the form \eqref{eqn:case_2_kraus_operator}, we use
\begin{equation}
    a_{m,\vec{k}}^{(\text{II})} = \sqrt{\frac{(1 - e^{-\alpha t})^m e^{-cg(t, \alpha)}}{\alpha^m m!}} \prod_{j=1}^m \sqrt{\gamma_{\vec{k}_j}}\norm{L_{\vec{k}_j}}.
\end{equation}

When the expectation values of a filtered observable $\hat{O}'$ are measured for each of the Kraus circuits $U_{K_{m,\vec{k}}}(t)$, the results must be recombined with weights $|a_{m,\vec{k}}(t)|^2$ on a classical computer to recover the time-dependent expectation value of the original system observable:
\begin{equation}
    \langle \hat{O}_{\text{sys}} \rangle_{\rho(t)} = \sum_{m}\sum_{\vec{k}} |\alpha_{m,\vec{k}}(t)|^2 \langle \hat{O}' \rangle_{U_{K_{m,\vec{k}}}}(t).
\end{equation}


\subsubsection{Representations of Systems Satisfying Commutation Relations (i)-(iv)}
\label{sec:representation_of_systems_i_iv}
So far, we have shown in a general way how the class of systems satisfying conditions 1-3 admit a Kraus series representation where each Kraus operator can be simulated with circuits of $t$-independent complexity. We will now focus on a subclass of these systems, which one can recognize immediately without analyzing the superoperator dynamics of $\mathcal{H}$ and $\mathcal{L}$. These are the systems which satisfy the commutation relations (i)-(iv) proposed in the introduction, which we will now prove satisfy conditions 1-3.

First, we will establish that a system satisfying the commutation relations (iii) and (iv) satisfies the commutation relation in condition 3. We expand $[\mathcal{H}, \mathcal{L}]$ in terms of $V_H$ and the $V_n$ operators defined in \eqref{eqn:V_H} and \eqref{eqn:V_n}, and we obtain
\begin{equation}
[\mathcal{H}, \mathcal{L}] = \sum_n \frac{-i}{\hbar}\left( [V_H, V_n]\otimes \overline{V_n} - V_n \otimes \overline{[V_H, V_n]} \right).
\label{eqn:comm_HL_expansion}
\end{equation}

After expanding $[V_H, V_n]$ and applying (iii) and (iv), we obtain
\begin{align}
    [V_H, V_n] &= \sum_{n'}\sqrt{\gamma_n}\left( [H, L_n] + \frac{-i\hbar\gamma_{n'}}{2}[L_{n'}^\dagger L_{n'}, L_n]\right) \\
    &= \sqrt{\gamma_n}\left(\nu L_n  + \frac{-i\hbar\lambda}{2} L_n\right) \\
    &= (\nu - i\hbar\lambda/2)V_n.
    \label{eqn:V_H_V_n_commutator}
\end{align}

Substituting \eqref{eqn:V_H_V_n_commutator} into \eqref{eqn:comm_HL_expansion} and simplifying yields the commutation relation 
\begin{equation}
    [\mathcal{H}, \mathcal{L}] = (2\Im[\nu]/\hbar - \Re[\lambda])\mathcal{L}
\end{equation}
which shows that the system satisfies condition 3 with $\alpha = 2\Im[\nu]/\hbar - \Re[\lambda] \ge 0$ and $c = 0$.\\

\begin{figure}
    \centering
    \includegraphics[width=\linewidth]{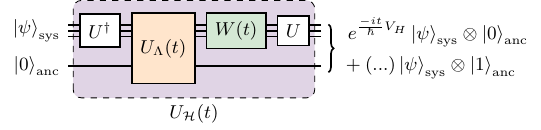}
    \caption{Quantum circuit diagram for computing $U_{\mathcal{H}}(t)$ when the commutation relations (i) and (ii) are satisfied and $\exp({\frac{-it}{\hbar}V_H})$ admits the decomposition \eqref{eqn:VH_decomposition}. Since $W(t)$ and $\Lambda(t)$ are diagonal, they can easily be represented by parameterized quantum circuits such as those shown in \figref{\ref{fig:diagonal_unitary_circuit}} and \figref{\ref{fig:diagonal_contraction_circuit}} respectively.}
    \label{fig:U_H_circuit}
\end{figure}

With regards to condition 2, we argue that as long as a system's Lindblad operators $L_n$ can be truncated to a bounded matrix form, then the $L_n$ can be re-scaled per \eqref{eqn:rescaled_lindblad_ops}, and can be subsequently realized as a dilated unitary operation. This satisfies condition 2. The only condition that remains to be shown is condition 1, namely that $e^{t\mathcal{H}}$ can be simulated using a circuit of $t$-independent gate complexity. Using \eqref{eqn:H_superop_dynamics}, we have already shown that it suffices to show that $\exp(-itV_H/\hbar)$ can be realized via the Sz.-Nagy dilation \eqref{eqn:U_H_sznagy_dilation} with $t$-independent gate complexity, which we will proceed to prove. From (i) and (ii) it immediately follows that $V_H$ commutes with its Hermitian adjoint ($[V_H, V_H^\dagger]= 0$) and hence is a normal operator. This means that $\exp(\frac{-it}{\hbar}V_H)$ can be diagonalized in the form
\begin{equation}
    e^{\frac{-it}{\hbar}V_H} = UW(t)\Lambda(t)U^\dagger,
    \label{eqn:VH_decomposition}
\end{equation}
where $W(t)$ and $\Lambda(t)$ are diagonal matrices with entries of the form $e^{-it\omega_i}$ and $e^{-t\lambda_i}$ respectively for real scalars $\omega_i, \lambda_i$. From the positive semi-definiteness of all $L_n^\dagger L_n$ terms in $V_H$, we also know that $\lambda_i \ge 0$ for all $i$ (which reflects the fact that $\exp({\frac{-it}{\hbar}V_H})$ is a contraction). Since $W(t)$ and $\Lambda(t)$ are diagonal, they can easily be implemented by parameterized quantum circuits. $W(t)$ is unitary, which means no ancilla qubits are required to realize it, whereas $\Lambda(t)$ must be implemented through a Sz.-Nagy dilation $U_{\Lambda(t)}$. Finally, the diagonalizing unitary $U$ can be implemented through a number of unitary decomposition methods such as the Quantum Shannon decomposition \cite{shende_synthesis_2006}, which can decompose an $N$-qubit unitary into a quantum circuit with at most $O(4^N)$ gates. We show the full quantum circuit diagram for realizing $U_{\mathcal{H}}(t)$ in \figref{\ref{fig:U_H_circuit}}.

\begin{figure}[h]
    \centering
    \includegraphics[width=\linewidth]{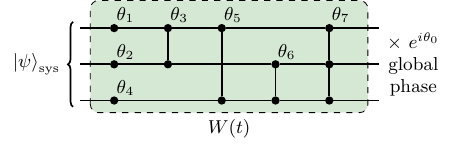}
    \caption{Quantum circuit diagram for an arbitrary $N=3$ qubit diagonal unitary operator $W(t)$ implemented through a sequence of controlled phase gates. Since the global phase factor $e^{i\theta_0}$ has no effect on measurement outcomes, the parameter $\theta_0$ is often ignored.}
    \label{fig:diagonal_unitary_circuit}
\end{figure}

Since the parameterized unitaries $W(t)$ and $U_{\Lambda}(t)$ have $2^{N}$ degrees of freedom for $N$ qubits, their circuit implementations must contain at least $2^N$ parameterized gates, ideally in such a manner that computing the gate parameters for a specific time $t$ is computationally efficient. In \figref{\ref{fig:diagonal_unitary_circuit}} we show a quantum circuit that implements $W(t)$ using a vector of $2^N$ real parameters $\vec{\theta} = (\theta_0~\theta_1~...~\theta_{2^N-1})^T$ corresponding to controlled phase gates, where the control qubits of each $\theta_n$ phase gate are the ``1" bits of the binary representation of $n$. The parameters $\vec{\theta}$ that realize a $W(t)$ with diagonal elements of the form $e^{-it\omega_n}$ can be computed according to the matrix equation
\begin{equation}
    \vec{\theta}(t) = tQ_N\vec{\omega}
    \label{eqn:u_mapping}
\end{equation}
where $\vec{\omega} = (\omega_0~\omega_1~ ...~ \omega_{2^N-1})^T$ and $Q_N$ is the $N$-qubit parameter transform matrix, given by:

\begin{equation}
    Q_N = \begin{pmatrix}
    1 & 0 \\ -1 & 1
    \end{pmatrix}^{\otimes N}.
    \label{eqn:Q_N_matrix}
\end{equation}

\begin{figure}[h!]
    \centering
    \includegraphics[width=\linewidth]{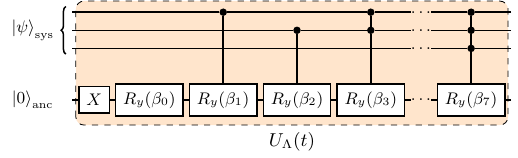}
    \caption{Quantum circuit diagram for the Sz.-Nagy dilation of an arbitrary 3-qubit diagonal contraction operator $\Lambda(t)$. It is implemented through a sequence of controlled $R_y$ gates on the ancilla qubit. Here, $R_y(\beta) \equiv \exp(\frac{-i\beta}{2}Y)$, where $Y$ is the Pauli $Y$ operator.}
    \label{fig:diagonal_contraction_circuit}
\end{figure}

In \figref{\ref{fig:diagonal_contraction_circuit}}, we show a similar quantum circuit that implements $U_{\Lambda}(t)$ by performing parameterized controlled $R_y$ gates on a single ancilla qubit, with parameter vector $\vec{\beta} = (\beta_0~\beta_1~ ...~ \beta_{2^N-1})^T$. In a manner similar to \eqref{eqn:u_mapping}, we can obtain the parameter vector $\vec{\beta}(t)$ corresponding to the diagonal elements $\vec{v}(t)$ of $\Lambda(t)$ by computing
\begin{equation}
    \vec{\beta}(t) = -2Q_N\arcsin(\vec{v}(t)).
    \label{eqn:v_mapping}
\end{equation}

The time-dependent parameter mappings of both $\vec{\theta}(t)$ in \eqref{eqn:u_mapping} and $\vec{\beta}(t)$ in \eqref{eqn:v_mapping} require a $2^N \times 2^N$ classical matrix multiplication by $Q_N$ to determine the parameters for $W(t)$ and $\Lambda(t)$. While it may appear that the complexity of this parameter mapping might dominate the complexity of simulating a large system on a quantum computer, it can be shown that $Q_N$ can be factored as a tensor power of a $2\times 2$ matrix, and thus the multiplication of $Q_N$ times a vector can in fact be carried out efficiently in $O(2^N\log(2^N))$ using a divide-and-conquer algorithm similar to the fast Walsh-Hadamard transform \cite{johnson_search_2000}. We give additional details regarding this parameter mapping process in \appref{\ref{sec:parameters_for_diagonal_quantum_circuits}}.

\subsection{Analysis of Accuracy and Time Complexity}
\label{sec:analysis_of_accuracy_and_time_complexity}

In this section, we combine the results of our previous analysis and derive bounds for time complexity, as well as provide some rough analysis of the error robustness of the proposed realization of systems satisfying conditions 1-3.

\subsubsection{Accuracy of Kraus Series Approximation}
\label{sec:accuracy_of_kraus_series_approximation}

First, we analyze the convergence of the general Kraus series representations for case (I) and case (II) in the general case of a system with $N_L$ Lindblad operators. We recall that the Kraus series (in superoperator form) is given by \eqref{eqn:lindblad_superoperator_kraus_series}. For both case (I) and case (II) systems, the terms corresponding to each $m$ in this series can be written in the form:
\begin{equation}
    \sum_{\vec{k} \in \{ 1, 2, ..., N_L\}^m} K_{m,\vec{k}}(t)^{\otimesbar 2} = e^{t\mathcal{H}}h(t) \frac{(f(t)\mathcal{L})^m}{m!}
    \label{eqn:kraus_term_factors}
\end{equation}
where $h(t)$ and $f(t)$ are both real non-negative functions of time. For case (I) and (II) systems, they take the form
\begin{align}
    h(t) &= \begin{cases}
        e^{-ct^2/2}, & \text{Case (I)} \\
        e^{cg(t,\alpha)}, & \text{Case (II)}
    \end{cases} \label{eqn:h_cases} \\
    f(t) &= \begin{cases}
        t, &  \text{Case (I)} \\
        (1-e^{-\alpha t})/\alpha, &  \text{Case (II)}
    \end{cases} \label{eqn:f_cases}.
\end{align}

Since we have assumed $\mathcal{L}$ is bounded and admits a sparse matrix representation that can be realized with a quantum circuit (a consequence of condition 2 being satisfied), then the Kraus series \eqref{eqn:kraus_term_factors} converges pointwise for every $t \ge 0$. Moreover if $f(t)$ and $h(t)$ are bounded, then the series also converges uniformly for $t \ge 0$. Each term in the right-hand side of \eqref{eqn:kraus_term_factors} corresponds to a term in a Taylor series approximation to the exponential of $\mathcal{L}$. In \appref{\ref{sec:bounds_on_kraus_series_error}}, we show that if the Kraus series is evaluated up to a finite order $m = M$ yielding an approximation $\vec{\rho}_M(t)$, the error with respect to the true time evolution $\vec{\rho}(t)$ can be bounded by
\begin{equation}
    \norm{\vec{\rho}(t) - \vec{\rho}_M(t)} \le \left( \norm{e^{t\mathcal{H}}}h(t) \eta(t) \right) \frac{(f(t)\hsnorm{\mathcal{L}})^{M+1}}{(M + 1)!},
    \label{eqn:kraus_series_error_bound}
\end{equation}
where $\eta(t)$ is given by
\begin{equation}
    \eta(t) = \left(1 - \frac{f(t)\hsnorm{\mathcal{L}}}{(M+1)}\right)^{-1},
    \label{eqn:kraus_series_error_bound_eta}
\end{equation}
and $\hsnorm{A} = \sqrt{\Tr(A^\dagger A)}$ denotes the Hilbert-Schmidt norm (also known as the Frobenius norm for matrices). It can be shown that $\hsnorm{\mathcal{L}}$ is bounded by
\begin{equation}
    \hsnorm{\mathcal{L}} = \hsnorm{\sum_n \gamma_n L_n^{\otimesbar 2}} \le \sum_n \gamma_n \hsnorm{L_n}^2
    \label{eqn:L_superop_bound}
\end{equation}
and that $\norm{e^{t\mathcal{H}}}$ is bounded by
\begin{equation}
    \norm{e^{t\mathcal{H}}} \le C e^{-t\lambda_L} \le C
    \label{eqn:exp_H_superop_bound}
\end{equation}
where $C$ is some constant and $\lambda_L$ is the smallest eigenvalue of $(\sum_{n} \gamma_n L_n^\dagger L_n)$. For systems satisfying conditions (i)-(iv), $C = 1$ due to the normality of $V_H$.

\begin{figure}
    \centering
    \includegraphics[width=\linewidth]{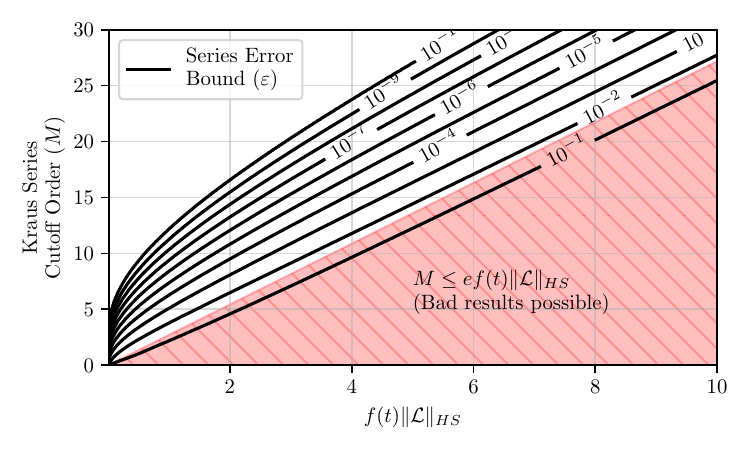}
    \caption{Contour plot of the series error bound $\varepsilon$ with respect to $M$ and $(f(t)\hsnorm{\mathcal{L}})$ for the ``worst case" values of $h(t) = \norm{e^{t\mathcal{H}}} = 1$. For small values of $f(t)\hsnorm{\mathcal{L}}$, $M$ exhibits almost-linear scaling of the form $M \propto (f(t)\hsnorm{\mathcal{L}})$, provided that $M$ at least satisfies \eqref{eqn:M_bound_1} and \eqref{eqn:M_bound_2} for a small $\delta$. If a cutoff order of $M < ef(t)\hsnorm{\mathcal{L}}$ is chosen (thereby violating \eqref{eqn:M_bound_1}), the truncated series is not guaranteed to give accurate results.}
    \label{fig:kraus_series_error_contours}
\end{figure}

Using \eqref{eqn:kraus_series_error_bound}, we can derive a rough lower bound on $M$ needed to guarantee that $\norm{\vec{\rho}(t) - \vec{\rho}_M(t)} \le \varepsilon$ for a desired error tolerance $\varepsilon$. In \appref{\ref{sec:bounds_on_kraus_series_error}
} we show that if $M$ satisfies the inequalities
\begin{equation}
    M \ge e (f(t)\hsnorm{\mathcal{L}})^{1 + \delta},
    \label{eqn:M_bound_1}
\end{equation}
(where $e \approx 2.72$) and
\begin{equation}
    M \ge \frac{\log(\norm{e^{t\mathcal{H}}}h(t)/\varepsilon)}{\delta \log(f(t)\hsnorm{\mathcal{L}})}
    \label{eqn:M_bound_2}    
\end{equation}
for any $\delta > 0$, then $\norm{\vec{\rho}(t) - \vec{\rho}_M(t)} \le \varepsilon$. In the limit of $\delta \ll 1$ we observe that the number of terms $M$ in the approximation scales almost linearly with $f(t)\hsnorm{\mathcal{L}}$. For a fixed $t$ and $\delta$, we observe logarithmic scaling with respect to $1/\varepsilon$, which is also desirable. In \figref{\ref{fig:kraus_series_error_contours}}, we plot the contours of the error bound $\varepsilon$ in \eqref{eqn:kraus_series_error_bound} for small values of $f(t)\hsnorm{\mathcal{L}}$ and verify that the approximately linear scaling of $M$ with respect to $f(t)\hsnorm{\mathcal{L}}$ holds in this regime.

For case (I) systems, the Kraus operators can be written in the form \eqref{eqn:kraus_term_factors} with $f(t) = t$ and $h(t) = e^{-ct^2/2}$. Since $f(t)$ is linear and unbounded, our analysis suggests that selecting $M \propto t$ is sufficient in the worst case to yield good asymptotic accuracy with a fixed $\varepsilon$; however if $c > 0$ or $\lambda_L > 0$ then one only needs to scale $M$ linearly with respect to $t$ until the point at which $Ce^{-\lambda_L t}h(t) \ll \varepsilon$, which gives the asymptotic bound $M \sim O(1)$. For case (II) systems, we have $f(t) = (1-e^{-\alpha t})/\alpha$ and $h(t) = e^{cg(t, \alpha)}$, which are both bounded by $1$. This means that the case (II) Kraus series converges uniformly and exponentially fast with respect to $t$ and it suffices to select a constant cutoff $M \sim O(1)$. 

To summarize, our analysis shows that for systems with $\alpha > 0$ or $c > 0$, the Kraus series converges quickly due to the rapid damping in the system, suggesting that the series cutoff $M$ is asymptotically independent of $t$. However, in the ``worst case" where $\alpha = c = \lambda_L = 0$, $M$ must be selected in a manner that scales almost linearly with $t$.

\subsubsection{Time Complexity (General Case)}

In the previous sections, we showed that each term in the Kraus series representation can be simulated by a circuit of the kind shown in \figref{\ref{fig:kraus_operator_circuit}}, which applies a sequence of $m$ re-scaled Lindblad operators (realized as the unitaries $U_{A_{\vec{k}_i}}$) followed by the effective Hamiltonian time evolution (realized as $U_{\mathcal{H}}$). Using standard unitary decomposition methods, we argued that each operator $L_n$ could be realized using $O(d^2)$ gates on an ideal quantum computer. Since NISQ devices are still far from ideal, we must justify that the $O(d^2)$ bound for unitary decomposition still holds under the constraints of NISQ quantum devices, such as a device having limited (but universal) native gate sets, or a device only allowing two-qubit interactions in accordance with the device's limited qubit topology. With regards to the first question, there is a substantial body of literature addressing it \cite{shende_synthesis_2006, kitaev_quantum_1997, dawson_solovay-kitaev_2005, reck_experimental_1994, barratt_parallel_2021, aharonov_power_2009, davis_towards_2020, kuperberg_breaking_2023}. Perhaps the most celebrated of these results is the generalized Solovay-Kitaev algorithm \cite{kitaev_quantum_1997, dawson_solovay-kitaev_2005}, which from any circuit of $n$ constant-qudit gates can construct a circuit with $O( n \log^c(n/\varepsilon))$ device-native gates up to approximation error $\varepsilon$. Recent improvements on this result have found a highly-efficient $c \le 1.441$ \cite{kuperberg_breaking_2023}. With regards to the second question of topology-aware decomposition of unitaries, there also exist many useful results \cite{davis_towards_2020,aharonov_power_2009,shende_synthesis_2006}. One analysis has shown that even in the most restrictive linear qubit topology, the two-qubit gate cost for realizing arbitrary unitaries increases by at most a constant factor of 9 \cite{shende_synthesis_2006}, thereby retaining the $O(d^2)$ asymptotic cost.

Now we will analyze the general time complexity of our method. If the effective Hamiltonian evolution $U_\mathcal{H}(t)$ can be realized with gate complexity $O(T_{\mathcal{H}}(d))$ in accordance with condition 1, we find that the total gate complexity of evaluating the entire Kraus series for $N_L$ Lindblad operators up to order $M$ is $ \sum_{m=0}^M (N_L)^m(O(T_H(d)) + m O(d^2))$ in the most general case. This is asymptotically equivalent to
\begin{equation}
    O((N_L)^M(T_{\mathcal{H}}(d) + Md^2)).
\end{equation}
The prefactor $(N_L)^M$ shows a rather unforgiving exponential growth with respect to the number of Lindblad operators, though we suggest that this prefactor can be managed for systems with few Lindblad operators or systems with a Kraus series that can be simplified using the methods in \secref{\ref{sec:simplifying_the_kraus_series}}. It is also worth commenting that the number of gates per circuit only grows as $O(T_{\mathcal{H}}(d) + M d^2)$, where we have shown the prefactor $M$ to be asymptotically $t$-independent except in the rare case where the quantities $\lambda_L$, $\alpha$, and $c$ are all zero.

\subsubsection{Time Complexity (Concrete Case)}

Next, we consider the concrete case where the commutation relations (i)-(iv) are satisfied, and we derive asymptotic bounds for the time complexity $T_{\mathcal{H}}(t)$ of simulating $e^{t\mathcal{H}}$.  For these systems, we found that $V_H$ is normal and admits the parameterized circuit representation shown in \figref{\ref{fig:U_H_circuit}}, where we proposed the circuit realizations of the diagonal unitaries $W(t)$ and $U_{\Lambda}(t)$ in \figref{\ref{fig:diagonal_unitary_circuit}} and \figref{\ref{fig:diagonal_contraction_circuit}} respectively. These circuits used a binary encoding scheme and both make use of the fast $N$-qubit parameter transformation $Q_N$ in \eqref{eqn:Q_N_matrix}, requiring \eqref{eqn:u_mapping} and \eqref{eqn:v_mapping} to be evaluated on a classical computer. These circuits use single-qubit unitaries with $m$ control bits, each of which can be decomposed into $O(m)$ one-qubit and two-qubit gates, following the methods in \cite{vale_decomposition_2023}. This results in an overall gate complexity of $\sum_{n=1}^{2^N} O(\log(n)) = O(2^N\log(2^N))$ for $N$-qubit diagonal operators. For systems in a Hilbert space of dimension $d$, this means that the circuits for $W(t)$ and $U_{\Lambda}(t)$ have $O(d\log(d))$ gate complexity. Since the transform $Q_N$ can also be computed in $O(2^N\log(2^N))$, the complexity of classically computing the time-dependent parameter mappings is also $O(d\log(d))$. Even though there exist more efficient representations of diagonal quantum operators requiring at most $O(d)$ gates \cite{bullock_smaller_2003}, these representations use Gray code encodings of the target bits instead of a standard binary encoding, which have parameter transform matrices that cannot be factored as the Kronecker product of smaller matrices (e.g. the matrix $\mathbf{\eta}^{\oplus}$ as introduced in \cite{bullock_smaller_2003}). In these methods, the classical computation of circuit parameters in $O(d^2)$ time dominates the $O(d)$ quantum gate complexity, which we consider to be less desirable than having a balanced $O(d \log(d))$ quantum and classical workload.

Since the sub-circuits for $W(t)$ and $U_{\Lambda}(t)$ in \figref{\ref{fig:U_H_circuit}} can both be realized with $O(d\log(d))$ quantum and classical complexity, the overall gate complexity of the system is limited by that of computing the unitary $U$ that diagonalizes $V_H$. In general, this can be accomplished with $O(d^2)$ gates through a unitary decomposition of $U$, unless the system and its Lindblad operators are already transformed into a diagonalized basis (which is common for many kinds of simulations). If this is the case, the entire $U_{\mathcal{H}}(t)$ can be computed rather efficiently with time complexity
\begin{equation}
    T_{\mathcal{H}}(t) \sim O(d\log(d))
\end{equation}
for both the quantum and classical workloads. Otherwise, we have $T_{\mathcal{H}}(t) \sim O(d^2)$ in the general case.

\section{Applications}
\label{sec:applications}


\subsection{The Multi-Qubit Continuous-Time Pauli Channel}

The multi-qubit continuous-time Pauli channel is an important model that describes the dynamics of unbiased decoherence and noise in degenerate quantum systems, such as quantum information processors \cite{harper_fast_2021, chen_learnability_2023} and coupled spin-1/2 particles \cite{bhattacharya_exact_2017}. The Pauli channel assumes a trivial Hamiltonian $H = 0$ and uses Lindblad operators consisting of $N$-qubit Pauli strings $\Pi_n$ of the form
\begin{equation}
    L_n = \Pi_n = (\sigma_{n1} \otimes \sigma_{n2} \otimes ... \otimes \sigma_{nN}),
    \label{eqn:pauli_strings}
\end{equation}
where each $\sigma_{ni}$ is one of the four single-qubit Pauli operators $\{ I,X,Y,Z \}$. One important feature of Pauli strings is the fact that the conjugation of any operator by any two Pauli strings is commutative. Specifically, this means that
\begin{equation}
    \Pi_{n}\Pi_{n'}\rho \Pi_{n'}^\dagger \Pi_{n}^\dagger = \Pi_{n'}\Pi_{n}\rho \Pi_{n}^\dagger \Pi_{n'}^\dagger
    \label{eqn:pauli_string_commutator}
\end{equation}
holds for any two Pauli strings $\Pi_n, \Pi_{n'}$ and any operator $\rho$. In addition, every Pauli string satisfies the identity $(\Pi_n^{\otimesbar 2})^2 = I$.
These two properties are important because they can be exploited to reduce the Kraus series representation of a multi-qubit continuous-time Pauli channel to a finite number of terms, as shown in \secref{\ref{sec:simplifying_the_kraus_series}}. A proof of \eqref{eqn:pauli_string_commutator} is given in \appref{\ref{sec:results_on_pauli_channels}}.

Since $H = 0$ and $\Pi_n^\dagger\Pi_n = I$ for all Pauli strings, the system satisfies (i)-(iv) with $\lambda = \sum_n \gamma_n$ and $\nu = 0$, so $[\mathcal{H},\mathcal{L}] = 0$, which results in a case (I) Kraus series representation. Although we could immediately write down the Kraus series terms in the form \eqref{eqn:case_1_kraus_operator}, we observe that \eqref{eqn:pauli_string_commutator} directly implies that $[V_n^{\otimesbar 2}, V_{n'}^{\otimesbar 2}] = 0$, which allows us to decompose the $e^{t\mathcal{L}}$ in \eqref{eqn:case_1_A_env} as the exponential of each Lindblad operator individually:

\begin{equation}
    e^{t\mathcal{L}} = \exp\left(t\sum_{n=1}^{N_L} V_n^{\otimesbar 2}\right) = \prod_{n=1}^{N_L} e^{t V_n^{\otimesbar 2}}.
\label{eqn:case_1_V_n_exponential_split}
\end{equation}

Since $(\Pi_n^{\otimesbar 2})^2 = I^{\otimesbar 2}$, we apply the hyperbolic identity $e^{t\Pi_n^{\otimesbar 2}} = I^{\otimesbar 2} \cosh(t) + \Pi_n^{\otimesbar 2}\sinh(t)$ and expand the right-hand side of \eqref{eqn:case_1_V_n_exponential_split} into
\begin{equation}
e^{t\mathcal{L}} = \prod_{n=1}^{N_L} [(\sqrt{\cosh(\gamma_n t)} I)^{\otimesbar 2} + (\sqrt{\sinh(\gamma_n t)} \Pi_n)^{\otimesbar 2}],
\end{equation}
which we then substitute back into \eqref{eqn:case_1_A_env} and simplify to obtain a finite Kraus series of the form
\begin{equation}
    e^{t(\mathcal{H} + \mathcal{L})} = \sum_{E \subseteq \{ 1, 2, ..., N_L\}} K_{E}(t)^{\otimesbar 2},
\end{equation}
where each Kraus operator $K_E$ is indexed by an ``error set" $E$ corresponding to one of the of the $2^N$ subsets of the Lindblad Pauli strings. The Kraus operators $K_E$ take the form
\begin{equation}
    K_{E}(t) = \sqrt{p_{E}(t)} \prod_{n \in E} \Pi_n,
    \label{eqn:pauli_channel_kraus_series}
\end{equation}
where $p_{E}(t)$ can be interpreted as the probability for the subset of ``Pauli errors" $\{\Pi_{n}\}_{n \in E}$ having occurred when the system's $N$ qubits are measured at a time $t$. This probability $p_{E}(t)$ is given by
\begin{equation}
    p_E(t) = \prod_{n=1}^{N_L} \frac{1 + (-1)^{\mathbb{I}_E(n)}e^{-\gamma_n t}}{2},
\end{equation}
where $\mathbb{I}_E(n) = 1$ when $n \in E$ and $0$ otherwise.

\begin{figure}[h]
    \centering
    \includegraphics[width=\linewidth]{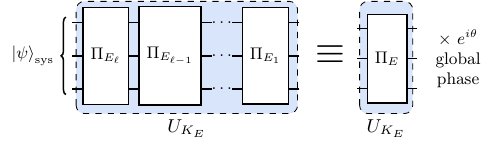}
    \caption{Kraus circuits for the multi-qubit continuous-time Pauli channel. Each sequence of Pauli strings indexed by the ``Pauli error" set $E$ can be reduced to a single Pauli string $\Pi_E$ up to a global phase.}
    \label{fig:pauli_channel_circuit}
\end{figure}

The Kraus operators in \eqref{eqn:pauli_channel_kraus_series} consist of time-dependent scalars $a_E(t) = \sqrt{p_E(t)}$ times products of Pauli strings. These products of Pauli strings are unitary and do not require a dilation to be realized efficiently on a quantum computer. In fact, by applying the product identities of Pauli operators, the product of Pauli strings $\prod_{n \in E} \Pi_{n}$ can be simplified to a single Pauli string $\Pi_E = \sigma_{E1} \otimes \sigma_{E2} \otimes ...\otimes \sigma_{EN}$ times a global phase of $e^{i\theta}$ which can be ignored. Each Kraus circuit $U_{K_E}$ is thus time-independent and can be realized as shown in \figref{\ref{fig:pauli_channel_circuit}}, where the gate $\Pi_E$ applies a Pauli gate $\sigma_{Ei}$ to each qubit.

The Kraus circuits in \figref{\ref{fig:pauli_channel_circuit}} require $O(\log(d))$ quantum gates, where $d = 2^N$ is the dimension of an $N$-qubit Pauli channel. The circuits are also parameterless, with all of the time-dependence of the circuit incorporated through the re-scaling constants $a_E(t) = \sqrt{p_E(t)}$. Remarkably, this means that it is possible to extrapolate the entire trajectory of observables to many different $t$ values (both forward and backward in time) using only the results of each Kraus circuit applied to the initial state $\rho(0)$. This feature can be exploited in the NISQ error mitigation technique of zero-noise extrapolation, where the result of a noisy quantum computation (modeled by a sparse Pauli channel \cite{harper_fast_2021}) can be evolved backward in time to suppress the effect of Markovian noise \cite{giurgica-tiron_digital_2020}.

\subsection{The Damped Quantum Harmonic Oscillator}

The damped quantum harmonic oscillator is a foundational system that appears in many application areas in quantum mechanics, such as quantum optics \cite{fabre_modes_2020} and quantum field theory \cite{peskin_introduction_1995}. This system is also being actively researched as a platform for fault-tolerant quantum information processing with bosonic codes \cite{mirhosseini_high-dimensional_2015, michael_new_2016}. A quantum harmonic oscillator has a Hamiltonian of the form
\begin{equation}
    H = \hbar\omega\left( \frac{1}{2} + \hat{N}\right),
\end{equation}
where $\hat{N} = \hat{a}^\dagger\hat{a}$ is the state number operator and $\hat{a}^\dagger, \hat{a}$ are the raising and lowering operators of the system. If damping is applied to this system, the single Lindblad operator $L_1 = \hat{a}$ is used with damping $\gamma_1$. One can easily verify that this system satisfies the commutation relations (i)-(iv) with scalars $\nu = \hbar\omega$ and $\lambda = -\gamma_1$. It then follows that conditions 1-3 are satisfied, with $[\mathcal{H},\mathcal{L}] = \gamma_1\mathcal{L}$, resulting in a Kraus series representation of the form \eqref{eqn:case_2_single_kraus_operator} with $\alpha = \gamma_1$ and $c = 0$. This simplifies to
\begin{equation}
    K_{m}(t) = e^{-t[\frac{\gamma_1}{2}\hat{N} + i\omega(\frac{1}{2} + \hat{N})]}\sqrt{\frac{\gamma_1^m (1 - e^{-\gamma_1 t})^m}{m!}}\hat{a}^m,
    \label{eqn:damped_qho_kraus_operators}
\end{equation}
which is consistent with prior derivations of a Kraus representation for this system \cite{liu_kraus_2004}.

Since $\hat{a}$ theoretically acts on an infinite number of modes of $H$, it is common to truncate $\hat{a}$ to act only on modes up to the highest initially occupied eigenstate $\ket{m_{\max}}$ in $\rho_{\text{sys}}$. This truncation makes $L_1 = \hat{a}$ representable as a $(m_{\max}+1)\times (m_{\max}+1)$ nilpotent matrix of index $m_{\text{max}}+1$ (i.e. $L_1^{m_{\text{max}}+1} = 0$):

\begin{equation}
    L_1 = \begin{pmatrix}
    0 & \sqrt{1} & 0 & 0 & \hdots & 0 \\
    0 & 0 & \sqrt{2} & 0 & \hdots & 0\\
    0 & 0 & 0 & \sqrt{3} & \hdots & 0 \\
    0 & 0 & 0 & 0 & \ddots & \vdots \\
    \vdots & \vdots & \vdots &  \vdots & \ddots & \sqrt{m_{\text{max}}}\\
    0 & 0 & 0 & 0 & \hdots & 0
    \end{pmatrix}.
    \label{eqn:qho_truncated_L1}
\end{equation}
This means that the Kraus series can be cut down to a manageable set of only $M = m_{\text{max}} + 1$ Kraus operators that need to be realized as quantum circuits. Here, we will assume $m_{\max} = 2^N - 1$, where $N$ is the desired number of system qubits.

\begin{figure}[h]
    \centering
    \includegraphics[width=0.85\linewidth]{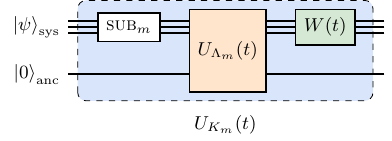}
    \caption{Kraus circuits for the damped quantum harmonic oscillator. These circuits fuse together the computation of $L_1^m$ with the effective Hamiltonian evolution circuit $U_{\mathcal{H}}(t)$ in \figref{\ref{fig:U_H_circuit}}.}
    \label{fig:qho_kraus_circuit}
\end{figure}

The Kraus operators $K_m(t)$ could be realized as circuits of the form shown in \figref{\ref{fig:kraus_operator_circuit}}, but this may be problematic because the repeated application of the unitary dilation $U_{A_1}$, where $A_1 = L_1/\norm{L_1}$ requires many quantum gates, which may induce noise. Fortunately, this gate cost can be reduced significantly by fusing the operations of the effective Hamiltonian evolution \eqref{eqn:VH_decomposition} and the evaluation of powers of the truncated $L_1$ operator in \eqref{eqn:qho_truncated_L1}. Specifically, we write
\begin{equation}
    K_m(t) = W(t)D_m(t)\textsc{sub}_m
    \label{eqn:dampled_qho_kraus_circuit_eqn}
\end{equation}
where $\textsc{sub}_m$ is the quantum circuit that implements subtraction by $m$ modulo $2^N$ on $N$ qubits, and $D_m(t)$ is the diagonal operator
\begin{equation}
    \begin{aligned}
    D_m(t) &= \sum_{n=0}^{2^N - (m+1)} \Big{(} e^{-n\gamma_1 t/2} \\
    & \times \sqrt{\frac{\gamma_1^m(1-e^{-\gamma_1 t})^m (n+m)!}{n!m!}} \ket{e_n}\bra{e_n}\Big{)},
    \end{aligned}
\end{equation}
with nonzero entries corresponding to the $n$-th super-diagonal of the matrix \eqref{eqn:qho_truncated_L1} raised to the power $n$. The operator $W(t) = e^{-it\omega(\frac{1}{2} + \hat{N})}$ is the time evolution operator for the oscillator Hamiltonian truncated to $2^N$ modes. To normalize the Kraus operators, we define $\Lambda_m(t) = D_m(t)/a_m(t)$ where $a_m(t) = \norm{D_m(t)}$ is the normalization coefficient. Then, the normalized Kraus operators can be realized as the circuit shown in \figref{\ref{fig:qho_kraus_circuit}}.

As shown previously, the gate $U_{\Lambda_m}(t)$ in \figref{\ref{fig:qho_kraus_circuit}} requires $O(d\log(d))$ quantum gates and $O(d\log(d))$ classical operations to compute the time-dependent gate parameters. However, $W(t)$ requires only $\log(d)$ parameterized gates, since it can be decomposed into only single-qubit phase gates of the form
\begin{equation}
    W(t) = e^{i\theta_0(t)}P(\theta_{2^0}(t)) \otimes P(\theta_{2^1}(t)) \otimes ... \otimes P(\theta_{2^N}(t))
\end{equation}
where $\theta_{2^n}(t) = 2^n\omega t$ and $\theta_0(t)$ is a global phase. Finally, the $\textsc{sub}_m$ gate can be implemented using at most $\log(m)+1$ $\textsc{sub}_1$ gates applied to subsets of the $N$ system qubits, each of which requires $O(N)$ gates with $N-1$ shared ancilla qubits \cite{li_class_2014}. Thus, the $\textsc{sub}_m$ gate can be decomposed into $O(\log(m)\log(d))$ elementary gates.

In total, the Kraus circuit in \figref{\ref{fig:qho_kraus_circuit}} requires $O(d\log(d))$ quantum gates and classical preprocessing operations, which is still more efficient than $O(d^2)$ for more general systems satisfying conditions (i)-(iv). In \tabref{\ref{tab:system_complexity_table}} we summarize the time complexity (quantum and classical) and number of Kraus circuits required for both this system, the continuous-time Pauli channel, and general systems satisfying conditions (i)-(iv).

\begin{table*}
    \centering
    \begin{tabular}{c  c@{\hskip 4mm} c@{\hskip 4mm} c}
    \hline\hline
    & MQCT Pauli Channel & Damped QHO & Systems Satisfying (i)-(iv) \\
    \hline
        Resources (per Kraus circuit): & & & \\
    \raggedright Quantum Gate Complexity & 
        $O(\log(d))$ & $O(d\log(d))$ & $O(d^2)$\\
    \raggedright Classical Pre-processing Complexity &
        $O(1)$ & $O(d\log(d))$ & $O(d\log(d))$\\
    \raggedright Number of Qubits & $\log_2(d)$ & $2\log_2(d)$ & $\log_2(d) + M + 1$\\[1mm]
    \hline
    \raggedright Number of Kraus circuits & 
         $2^{N_L}$ & $m_{\max} + 1 \le d$ & $\sum_{m=0}^M (N_L)^m < \infty$ \\[1mm]
    \hline\hline
    \end{tabular}
    \caption{Table of Kraus circuit resource requirements (quantum gate complexity, classical pre-processing complexity, and number of qubits) with respect to system dimension $d$. Results are shown for the multi-qubit continuous-time Pauli channel, the damped quantum harmonic oscillator, and for general systems satisfying the concrete conditions (i)-(iv) introduced in \secref{\ref{sec:introduction}}. For each system, bounds on the number of Kraus circuits are also shown.}
    \label{tab:system_complexity_table}
\end{table*}

\section{Conclusion}

In the beginning of this paper we introduced the superoperators $\mathcal{H}$ and $\mathcal{L}$ which can be used to describe the dynamics of open quantum systems under the Born-Markov approximation. For certain systems satisfying $[\mathcal{H},\mathcal{L}] = \alpha\mathcal{L} + c$ for $\alpha, c \geq 0$, we showed that a Kraus series representation of the system could be obtained, where each Kraus term in the series can be realized as a quantum circuit with bounded time-independent gate complexity. After analyzing the convergence of this Kraus series, we showed that the series can be truncated to a constant number of terms, which can be efficiently evaluated on NISQ quantum devices. To demonstrate how our framework can be applied, we analyzed two canonical open quantum systems, the multi-qubit Pauli channel and the damped quantum harmonic oscillator, and showed that they satisfy the concrete conditions (i)-(iv), and thus can be simulated efficiently with a series of Kraus circuits with $t$-independent depth. 

We anticipate that this framework may be useful for not only simulating open quantum systems (such as the two example systems discussed here), but also as a means of quickly preparing states that lie along the trajectories of these systems. The fact that our method distributes the complexity of state preparation across several circuits of bounded depth may yield a critical advantage when run on NISQ devices, since it minimizes state preparation error and incurs a low per-circuit gate complexity overhead compared to Trotter-based methods.

\subsubsection{Future Work}

Although the framework we have developed in this paper is promising for simulating the class of systems that satisfy conditions 1-3, it does not admit a straightforward generalization to arbitrary Markovian open quantum systems governed by the Lindblad equation. Our method exploits the fact that the commutation relations between $\mathcal{H}$ and $\mathcal{L}$ are such that the Zassenhaus expansion \eqref{eqn:zassenhaus} of the system dynamics collapses to a tractable form, allowing for the environment action superoperator $\mathcal{A}_{\text{env}}(t)$ to admit a series expansion. The discovery of  different or more general sets of conditions that facilitate closed-form time evolution formulas of open quantum systems may be a promising direction of future research. In addition, the discovery of new applications or NISQ hardware-specific implementations of systems satisfying conditions 1-3 may also prove a fruitful area for further investigation. This framework is presently being applied in ongoing computations with NISQ quantum hardware.

\begin{acknowledgements}
The authors would like to thank Dr. Gerald Cleaver of the Department of Physics at Baylor University for his insightful discussion and proofreading of the manuscript.
\end{acknowledgements}

\appendix

\section{Parameters for Diagonal Operator Quantum Circuits}
\label{sec:parameters_for_diagonal_quantum_circuits}

In this appendix section we provide additional details regarding the parameter mappings for the quantum circuit implementations of $W(t)$ (\figref{\ref{fig:diagonal_unitary_circuit}}) and $\Lambda(t)$ (\figref{\ref{fig:diagonal_contraction_circuit}}). We will also derive equations \eqref{eqn:u_mapping}-\eqref{eqn:v_mapping} and show how the transformation $Q_N$ in \eqref{eqn:Q_N_matrix} can be computed efficiently.

\subsection{Parameter mappings for the diagonal unitary $W(t)$:}

First, we consider the representation of an arbitrary $N$-qubit diagonal unitary $W(t) = \sum_{n=0}^{2^N-1} e^{-it\omega_n}\ket{e_n}\bra{e_n}$ using the circuit in \figref{\ref{fig:diagonal_unitary_circuit}} with parameter vector $\vec{\theta} = (\theta_0~\theta_1~...~\theta_{2^N-1})$. Adding up the phases accumulated by each diagonal state $\ket{e_n}$ from the circuit $W(t)$, we see that the parameters $\vec{\theta}(t)$ must be chosen such that:
\begin{equation}
    \exp\left(i\theta_0(t) + i\sum_{B_{s} \subseteq B_n} \theta_s(t) \right)\ket{e_n} = e^{it\omega_n}\ket{e_n}
    \label{eqn:W_phase_sum}
\end{equation}
where $B_n = \{ m_0, m_1, ... \}$ denotes the set of non-negative indices $m$ of the ``1" bits in the binary representation of $n$ (i.e. $B_n$ uniquely satisfies $\sum_{m \in B_n} 2^m = n$). Since $\theta_0$ corresponds to the global phase induced by $W(t)$, this parameter can be ignored in most situations, as it has no effect on measurement outcomes. Equating the left hand exponential to the right hand exponential in \eqref{eqn:W_phase_sum}, we obtain the linear system:

\begin{equation}
    \Bigg \lbrace \theta_0(t) + \sum_{B_s \subseteq B_n} \theta_s(t) = t\omega_n \quad (n = 0, 1, ..., 2^{N}-1)
\end{equation}

This system can be written in the matrix form:
\begin{equation}
    S_N \vec{\theta}(t) = t\vec{\omega}
\end{equation}
where the matrix $S_N$ factors as an order $N$ tensor power:
\begin{equation}
    S_N = \begin{pmatrix}
        1 & 0 \\
        1 & 1
    \end{pmatrix}^{\otimes N}
\end{equation}
Solving for $\vec{\theta}(t)$, we get:
\begin{equation}
    \vec{\theta}(t) = S_N^{-1} t\vec{\omega}
\end{equation}
where
\begin{equation}
    S_n^{-1} = \left[\begin{pmatrix}
        1 & 0 \\
        1 & 1
    \end{pmatrix}^{-1}\right]^{\otimes N} = \begin{pmatrix} 1 & 0 \\ -1 & 1\end{pmatrix}^{\otimes N} = Q_N
\end{equation}
from which we immediately obtain \eqref{eqn:u_mapping} and \eqref{eqn:Q_N_matrix}.

\subsection{Parameter mappings for the diagonal contraction operator $\Lambda(t)$:}

Next, we will consider the representation of the circuit $U_{\Lambda}(t)$ corresponding to an arbitrary diagonal contraction operator $\Lambda(t) = \sum_{n=0}^{2^N-1} \vec{v}_n(t)\ket{e_n}\bra{e_n}$, where $\norm{\vec{v}(t)} \leq 1$ for all $t \geq 0$. To realize the Sz.-Nagy dilation of $\Lambda(t)$, we proposed the circuit in \figref{\ref{fig:diagonal_contraction_circuit}}, which consists of a sequence of controlled $\hat{R}_y$ gates applied to a single ancilla qubit prepared in the $\ket{1}_{\text{anc}}$ state. The amplitude of a filtered basis state $\ket{e_n} \otimes \ket{0}_{\text{anc}}$ after applying this realization of  $U_{\Lambda}$ with parameter vector $\vec{\beta}$ is given by:
\begin{equation}
\begin{aligned}
    &(\bra{e_n}\otimes\bra{0}_{\text{anc}})U_{\Lambda}(t)(\ket{e_n}\otimes\ket{0}_{\text{anc}}) \\
    &\qquad\qquad = \bra{0}_{\text{anc}}e^{\frac{-i}{2}\left( \beta_0 + \sum_{B_s \subseteq B_n} \beta_s\right)\hat{Y}}\ket{1}_{\text{anc}} \\
    &\qquad\qquad = -\sin\left[\frac{1}{2}\left( \beta_0 + \sum_{B_s \subseteq B_n} \beta_s\right)\right]
\end{aligned}
\end{equation}
In order to represent a contraction operator with diagonal elements $\vec{v}(t)$, the parameters $\vec{\beta}(t)$ must then be selected such that they satisfy the linear system:
\begin{equation}
\begin{cases}
   \ \beta_0(t) + \sum_{B_s \subseteq B_n} \beta_s(t) = -2\arcsin(v_n(t)) \\[2mm] \quad  (n = 0, 1, ..., 2^{N}-1)
\end{cases} 
\end{equation}
Following the same procedure as with $W(t)$, the solution to this system can be written in matrix form:
\begin{equation}
    \vec{\beta}(t) = -2 S_N^{-1}\arcsin(\vec{v}(t)) = -2 Q_N\arcsin(\vec{v}(t))
\end{equation}
This agrees with the parameter mapping given in \eqref{eqn:v_mapping}.

\subsection{Computing the parameter transform $Q_N$ efficiently}

Above, we claimed that the product of $Q_N$ and an arbitrary vector could be computed efficiently on a classical computer in $O(d\log(d))$ time, where $d = 2^N$ for an $N$-qubit Hilbert space. This is achieved by a modified form of the fast Walsh-Hadamard transform, as described in \cite{johnson_search_2000}. Computing the product of $Q_{N}$ and a vector $\vec{u} = \vec{u}^{(0)}$ can be reduced to an iterative $N$-step procedure, where the $N$-th step intermediate result $\vec{u}^{(N)}$ contains the result of the product
\begin{equation}
     \vec{u}^{(N)} = Q_N\vec{u}^{(0)}.
\end{equation}

Since $Q_N$ can be factored as an order $N$ tensor power, we can apply the matrix $Q_N$ by sequentially applying the component of $Q_N$ corresponding to each qubit in the Hilbert space. This gives us the recurrent formula
\begin{equation}
    \vec{u}_n^{(t+1)} = \left(I^{\otimes(N-(t+1))} \otimes \begin{pmatrix}
        1 & 0 \\ -1 & 1 
    \end{pmatrix} \otimes I^{\otimes t} \right) \vec{u}^{(t)}.
\end{equation}
It follows that each entry $\vec{u}_n^{(t)}$ in $\vec{u}^{(t)}$ (for $n = 0, 1, ..., 2^N$) can be updated according to the procedure
\begin{equation}
    \vec{u}^{(t+1)}_{n} = \begin{cases}
        \vec{u}^{(t)}_{n}, & t \notin B_n\\
        \vec{u}^{(t)}_{n} - \vec{u}^{(t)}_{n-2^t}, & t \in B_n\\
    \end{cases}.
    \label{eqn:Q_n_recurrence_formula}
\end{equation}

In total, applying \eqref{eqn:Q_n_recurrence_formula} for $N$ steps requires $N \times 2^{N-1}$ subtraction operations, which for a Hilbert space of dimension $d = 2^N$ corresponds to a time complexity of $O(d\log(d))$.

\section{Bounds on Kraus Series Error}
\label{sec:bounds_on_kraus_series_error}

In this appendix section, we derive equations \eqref{eqn:kraus_series_error_bound}-\eqref{eqn:M_bound_2} by using properties of the Hilbert-Schmidt norm to bound the Kraus series error. 

\subsection{Error of the Kraus Series}

For a complete orthonormal basis $\ket{e_i}$ in some Hilbert space, the Hilbert-Schmidt norm is given by:
\begin{equation}
\hsnorm{A} = \sqrt{\Tr(A^\dagger A)} = \sqrt{\sum_{i,j} |\bra{e_i} A \ket{e_j}|^2}
\end{equation}

Now, let $E_M = \norm{\vec{\rho}(t) - \vec{\rho}_M(t)}$ be the error of a Kraus series evaluated up to order $m = M$. Expanding $E_M$ with the Hilbert Schmidt norm and using \eqref{eqn:lindblad_superoperator_kraus_series} and \eqref{eqn:kraus_term_factors}, we bound $E_M$ as follows:
\begin{align}
    E_M &= \norm{\vec{\rho}(t) - \vec{\rho}_M(t)} \notag \\
    &\le \norm{ e^{t\mathcal{H}}h(t)e^{f(t)\mathcal{L}}\vec{\rho}(0) - \sum_{m=0}^M e^{t\mathcal{H}}h(t)\frac{(f(t)\mathcal{L})^m}{m!} 
 \vec{\rho}(0)} \notag \\
    &\le \hsnorm{ e^{t\mathcal{H}}h(t)\left(e^{f(t)\mathcal{L}} - \sum_{m=0}^M \frac{(f(t)\mathcal{L})^m}{m!} \right)} \norm{\vec{\rho}(0)} \notag \\
    & \le \norm{e^{t\mathcal{H}}}h(t) \hsnorm{\sum_{m=(M+1)}^\infty \frac{(f(t)\mathcal{L})^m}{m!}} \notag \\
    & \le \norm{e^{t\mathcal{H}}} h(t) \sum_{m=(M+1)}^{\infty} \frac{(f(t)\hsnorm{\mathcal{L}})^m}{m!} \notag \\
    & \le \norm{e^{t\mathcal{H}}} h(t) \left[ \sum_{m=0}^{\infty} \left( \frac{f(t)\hsnorm{\mathcal{L}}}{M+1} \right)^m \right] \frac{(f(t)\hsnorm{\mathcal{L}})^{M+1}}{(M+1)!}. \label{eqn:E_M_series}
\end{align}

For $M > f(t)\hsnorm{\mathcal{L}}$, we expand the geometric series in \eqref{eqn:E_M_series} and re-arrange terms to obtain
\begin{equation}
    \frac{E_M}{\norm{e^{t\mathcal{H}}}h(t)} \le \left(\frac{1}{1 - \frac{f(t)\hsnorm{\mathcal{L}}}{(M+1)}}\right)\frac{(f(t)\hsnorm{\mathcal{L}})^{M+1}}{(M + 1)!},
\end{equation}
from which \eqref{eqn:kraus_series_error_bound} and \eqref{eqn:kraus_series_error_bound_eta} immediately follow. 
From Stirling's inequality, we have
\begin{equation}
    2\left( \frac{M}{e}\right)^{M+1} \le \sqrt{2\pi(M+1)}\left(\frac{(M+1)}{e}\right)^{(M+1)} \le (M+1)!, \label{eqn:stirling_inequality}
\end{equation}
which means that if $M$ is chosen such that \eqref{eqn:M_bound_1} is satisfied, then we obtain
\begin{align}
    \frac{E_M}{\norm{e^{t\mathcal{H}}} h(t)} &\leq \left(\frac{1}{1 - (1/2)}\right)\frac{(f(t)\hsnorm{\mathcal{L}})^{M+1}}{2(M/e)^{M+1}} \notag \\
    &\leq \frac{(f(t)\hsnorm{\mathcal{L}})^{M+1}}{(f(t)\hsnorm{\mathcal{L}})^{(1 + \delta)(M+1)}} \notag \\
    &\leq (f(t)\hsnorm{\mathcal{L}})^{-\delta(M+1)}. \label{eqn:log_bound}
\end{align}
After re-arranging \eqref{eqn:log_bound}, we obtain the following bound of $E_M$ for a desired error tolerance $\varepsilon$:
\begin{equation}
    E_M \le \frac{h(t)\norm{e^{t\mathcal{H}}}}{(f(t)\hsnorm{\mathcal{L}})^{\delta(M+1)}} \le \varepsilon. \label{eqn:error_tolerance_bound}
\end{equation}

Solving for $M$ on the right hand side in terms of $\varepsilon$ and the remaining factors, we obtain 
\begin{equation}
    \frac{\log(h(t)\hsnorm{e^{t\mathcal{H}}}/\varepsilon)}{\delta\log(f(t)\hsnorm{\mathcal{L}})} \leq M+1
\end{equation}

It immediately follows that if the second inequality \eqref{eqn:M_bound_2} is satisfied, then the error bound \eqref{eqn:error_tolerance_bound} holds.

\section{Results on Pauli Channels}
\label{sec:results_on_pauli_channels}

In this section, we give additional discussion regarding the continuous-time multi-qubit Pauli channel.

\subsection{Commutators of Pauli strings}

Here, we prove that \eqref{eqn:pauli_string_commutator} holds by deriving the commutator formula for arbitrary Pauli strings $\Pi_n, \Pi_{n'}$. Writing \eqref{eqn:pauli_string_commutator} in superoperator form, we obtain
\begin{equation}
    \Pi_n^{\otimesbar 2} \Pi_{n'}^{\otimesbar 2} \vec{\rho} = \Pi_{n'}^{\otimesbar 2} \Pi_{n}^{\otimesbar 2} \vec{\rho}
\end{equation}
for an arbitrary vector $\vec{\rho}$. Thus, proving that \eqref{eqn:pauli_string_commutator} holds is equivalent to proving that $[ \Pi_n^{\otimesbar 2}, \Pi_{n'}^{\otimesbar 2} ] = 0$. First, we consider the commutator and anti-commutator of two arbitrary Pauli strings $\Pi_n, \Pi_{n'}$, both of which can be expanded as the Kronecker product of commutators and anti-commutators of the Pauli operators that comprise them. For the commutator, we observe that
\begin{equation}
\begin{aligned}
    [\Pi_n, \Pi_{n'}] &= [\Pi_n^{(1)}, \Pi_{n'}^{(1)}] \\
    &=\frac{1}{2}[ \sigma_{n1}, \sigma_{n'1}] \otimes \{ \Pi_{n}^{(2)}, \Pi_{n'}^{(2)} \} \\
    &\quad + \frac{1}{2}\{ \sigma_{n1}, \sigma_{n'1} \} \otimes [ \Pi_{n}^{(2)}, \Pi_{n'}^{(2)} ]
\end{aligned},
\label{eqn:pauli_string_recursive_commutator}
\end{equation}
where we use $\Pi_{n}^{(j)} = \sigma_{nj} \otimes \sigma_{n(j+1)} \otimes ... \otimes \sigma_{nN}$ to denote the Pauli sub-string of $\Pi_{n}$ starting at the $j$th operator. Likewise for the anti-commutator, we have
\begin{equation}
\begin{aligned}
    \{\Pi_n, \Pi_{n'}\} &= \{\Pi_n^{(1)}, \Pi_{n'}^{(1)}\} \\
    &=\frac{1}{2}\{ \sigma_{n1}, \sigma_{n'1} \} \otimes \{ \Pi_{n}^{(2)}, \Pi_{n'}^{(2)} \} \\
    &\quad + \frac{1}{2}[ \sigma_{n1}, \sigma_{n'1} ] \otimes [ \Pi_{n}^{(2)}, \Pi_{n'}^{(2)} ].
\end{aligned}
\label{eqn:pauli_string_recursive_anticommutator}
\end{equation}
Since either $[ \sigma_i, \sigma_j ] = 0$ or $\{ \sigma_i, \sigma_j \} = 0$ for all $\sigma_i, \sigma_j \in \{I, X, Y, Z\}$, exactly one of the two terms in \eqref{eqn:pauli_string_recursive_commutator} and \eqref{eqn:pauli_string_recursive_anticommutator} must vanish. Thus, if we apply \eqref{eqn:pauli_string_recursive_commutator} and \eqref{eqn:pauli_string_recursive_anticommutator} recursively to expand $[ \Pi_n, \Pi_n']$ as the Kronecker product of commutators and anti-commutators of the corresponding Pauli operators, we obtain the formula
\begin{equation}
    [ \Pi_n, \Pi_n'] = \left(1 - \prod_{j=1}^N \tau_j\right)(S_1 \otimes S_2 \otimes ... \otimes S_N),
    \label{eqn:pauli_string_commutator_formula}
\end{equation}
where
\begin{equation}
    \tau_j = \begin{cases}
        +1, & [\sigma_{nj}, \sigma_{n'j}] = 0\\
        -1, & [\sigma_{nj}, \sigma_{n'j}] \neq 0
    \end{cases}
    \label{eqn:tau_j}
\end{equation}
and
\begin{equation}
    S_j = \begin{cases}
        \frac{1}{2}\{\sigma_{nj}, \sigma_{n'j}\}, & [\sigma_{nj}, \sigma_{n'j}] = 0\\
        \frac{1}{2}[\sigma_{nj}, \sigma_{n'j}], & [\sigma_{nj}, \sigma_{n'j}] \neq 0
    \end{cases}.
    \label{eqn:S_j}
\end{equation}
From inspecting \eqref{eqn:pauli_string_commutator_formula} and \eqref{eqn:tau_j}, we deduce that $[ \Pi_n, \Pi_{n'}] = 0$ if and only if the number of non-commuting pairs of Pauli operators $\sigma_{nj}, \sigma_{n'j}$ is even.

Finally, we note that the conjugated Kronecker product $\Pi_n^{\otimesbar 2}$ is itself a Pauli string with an added sign of $\pm 1$, since the map $\sigma_i \mapsto \overline{\sigma_i}$ fixes all Pauli operators except for ${Y} \mapsto \overline{Y} = -Y$. For an $N$-qubit string, we write
\begin{equation}
\Pi_n^{\otimesbar 2} = \pm(\sigma_{n1} \otimes \sigma_{n2} \otimes ... \otimes \sigma_{nN}) \otimes (\sigma_{n1} \otimes ... \otimes \sigma_{nN}).
\end{equation}
For any two such strings $\Pi_n^{\otimesbar 2}, \Pi_{n'}^{\otimesbar 2}$ every non-commuting Pauli operator pair $\sigma_{nj}, \sigma_{n'j}$ appears twice in the string, so there are an even number of these pairs. It follows from \eqref{eqn:pauli_string_commutator_formula} that $[\Pi_n^{\otimesbar 2}, \Pi_{n'}^{\otimesbar 2}] = 0$ as desired.

\medskip

%
\bibliography{main}

\end{document}